\documentclass[prl,twocolumn,superscriptaddress,showpacs,amsmath,amssymb]{revtex4}
\usepackage{graphicx}
\usepackage{indentfirst}
\usepackage{psfrag}
\usepackage{epsfig}
\usepackage{amsmath}
\usepackage{amssymb}
\usepackage{bm}
\usepackage{mathrsfs}
\newcommand{\be}{\begin{eqnarray}}
\newcommand{\ee}{\end{eqnarray}}

\newcommand{\hel}{\mathscr{ H}}
\newcommand{\tens}{\mathscr{ N}}
\newcommand{\tensq}{\mathscr{ T}}
\newcommand{\leh}{\mathscr{ L}}
\newcommand{\flow} {\mathscr{ F}}
\newcommand{\perd}{\mathscr{ P}}
\newcommand{\perdr}{\mathscr{ R}}
\newcommand{\qui}{\mathscr{ K}}
\newcommand{\s}{\mathscr{ S}}

\newcommand{\MM}{\mathscr{ M}}
\begin{document}

\title{The optical torque:  Electromagnetic spin and orbital angular  momenta conservation laws and their significance} 
\author{Manuel Nieto-Vesperinas }
\affiliation{Instituto de Ciencia de Materiales de Madrid, Consejo Superior de
Investigaciones Cient\'{i}ficas\\
 Campus de Cantoblanco, Madrid 28049, Spain.\\ www.icmm.csic.es/mnv; 
mnieto@icmm.csic.es }



\begin{abstract}
The physics involved in the fundamental conservation equations of the spin and orbital angular momenta leads to new laws and phenomena that are disclosed here. To this end, we analyse the scattering of an electromagnetic wavefield by the canonical system constituted by a small particle, which is assumed dipolar in the wide sense. Specifically, under  quite general conditions these laws lead to understanding   the contribution and weight of each of those angular momenta to the electromagnetic torque exerted by the field on the object, which is shown to consist  of  an extinction and a scattering, or recoil,  part. This leads to an interpretation of its effect  different to that taken up till now by many theoretical and experimental works, and implies that a part of the recoil torque cancels the usually called intrinsic torque which was often considered responsible of the particle spinning. In addition, we obtain the contribution of the spatial structure of the wave to this torque,  unknown  to this date,  showing its effect in the orbiting of the object, and demonstrating that it  often leads to a negative torque on a single particle, i.e. opposite to the incident helicity, producing an orbital motion contrary to its spinning.  Furthermore, we establish a decomposition of the electromagnetic torque into conservative and non-conservative components in which the helicity and the spin angular momentum play a role analogous to the energy and its flux  for electromagnetic forces. These phenomena are illustrated with examples of beams, also showing the difficulties of some paraxial formulations whose fields do not hold the transversality condition.
\end{abstract}
\maketitle

 \section{Introduction}
There is  a growing interest in phenomena related to the angular momenta of light and other electromagnetic fields. Advances in particle manipulation are making experiments,  their assessment, interpretation with theories, and applications of  increasing study and capability. This has opened a new area of research related to the twisting of both the polarization and the wavefronts of electomagnetic fields \cite{allenlibro,allenlibro1,babiker,yao,mansu,barnett,bliokh2,bliokh1,marston1,dola2,lipkin,alex,
cameron1,cameron2,philb,tang1,tang2,marru,molina1,molina2}, and their effects on matter \cite{beth,garces,
grier,dunlop2,dunlop3,dunlop4,laporta,dunlop,chaumet2,dogariu,cana0,cana,beckeva,mars,dog,brasse,chen1}

On the one hand, conservation laws for the helicity and spin of wavefields \cite{alex,cameron1,cameron2,philb,tang1,bliokh2} were recently established based on initial extensions \cite{lipkin} of known conservation equations of electromagnetism. Such laws appear as fundamental as those for the energy, and linear and angular momenta. In particular \cite{bliokh2} put forward the conservation of spin  and  orbital  angular momenta, separately. On the other hand, studies on the momenta transfer to matter, and their mechanical action on objects, are enlarging the area of optical manipulation of objects. Among these, new bizarre effects like negative optical torques (i.e.  opposite to the helicity of the illumination) on sets of particles have been predicted  \cite{chen1, nieto2015}  with circularly polarized plane waves, as well experimentally observed with Gaussian beams  on extended objects \cite{brasse}. This phenomenon keeps an analogy with pulling forces \cite{chen,novits,dogariu1}   that so much attention have recently attracted.

However, theory and experiments on electromagnetic torques are less developed and understood than their counterparts of optical and electromagnetic forces \cite{Ashkin1,Ashkin2,chen,novits,dogariu1,mazilu,chaumet,grier,neu,arias,quid1,quid2,albala,berry,chaumet3,MNV2010}. Observations are more difficult to control and to quantitatively interpret with existing  models.  With few exceptions \cite{mars,chen1}, most experimental \cite{laporta,dunlop} and theoretical   \cite{chaumet2,dogariu,cana0,cana,beckeva,mars,chaumet3} studies  employ a static formulation, (perhaps following the path of pioneering work \cite{beth}), which was shown \cite{nieto2015} to be  incomplete and not compatible with energy and angular momentum conservation.  Only for extremely small (i.e. Rayleigh) particles in terms of the wavelength the static approximation is valid; this is further discussed in Section 10, where the total electromagnetic torque is addressed.

Although several reports \cite{garces,grier,dunlop2,dunlop3,dunlop4,dunlop,mars} realize the need of energy absorption by the object in order that it experiences a torque, no explicit demonstration exists of the role played in this effect by the variation of incident spin and orbital angular momenta, even though  they are calculated in some cases after scattering \cite{bliokh1,marston1,mars,dog}. Moreover, studies based on the static approximation  \cite{laporta,dunlop,chaumet2,dogariu, cana,chaumet3}, only deal with the so-called intrinsic torque which, as as shown below cannot account for the angular momentum transfer nor can describe the resulting torque experienced by the object through energy absorption.

In a recent work \cite{nietoPRA1} we established the significance of the conservation of electromagnetic helicity for scattering objects and its relevance for energy transfer between small particles and molecules, as well as for  circular dichroism. Now the aim of this paper is two-fold: On the one hand, we integrate the conservation laws for the spin and orbital angular momenta, extracting their physical meaning for the interaction of fields and objects, specifically in the scattering of a wavefield by a small particle. On the other hand, we predict the contributions of each of these angular momenta conservation laws to the optical torque exerted by the field on the object; also taking into account that part due to the spatial structure of the field. This is illustrated with a  generally magnetodielectric bi-isotropic \cite{kong}  dipolar particle  in the wide sense, i.e. whose scattering is fully described by the first order partial waves, (namely, Mie coefficients if it were a sphere \cite{MNV2010,Nieto2011, chanlat}). This analysis opens a collection of new phenomena related to the electromagnetic torque, suceptible of further investigation from the theoretical and experimental points of view.
Our discussion is classical, however if one  considers the quantum nature of light, these results  may be obtained in terms of the number operators associated to left and right circular polarized modes, expressed by the product of the corresponding annihilation and creation photon operators, (see \cite{andrews1,andrews2,andrews3}), specially since methods  to identify twisted beams with different topological charge have been devised  \cite{andrews3,pu, boyd2}.

A Coulomb gauge is chosen dealing with fields that satisfy the transversality condition. This allows  a separation of the spin and the orbital angular momenta, thus confering sense to the existence of their respective  conservation laws \cite{cameron1, bliokh2}, separately. It is known, however,  that certain formulations of optical beams  have a problem with the transversality condition, and this is also noticed in this work.  Hence, within this context, and within the validity of these conservation laws, we  reach the following results:

1. The conservation laws for the spin and orbital angular momenta lead to their respective  contributions  to the time-averaged electromagnetic torque on the object, which from each of them  is composed of two parts: (a) A {\it extinction torque} due to the extraction of  the  corresponding (spin or orbital) angular momentum from the incident wave by scattering from the body, described by the interference between the incident and the scattered fields. (b) A {\it scattering} or {\it recoil torque} due to the scattered field interfering with itself. 

2. The recoil torque due to the orbital angular momentum conservation is equal to that yielded by the spin angular momentum conservation; both of them add to produce the total recoil torque, which is the same as that derived from the conservation of total angular momentum \cite{nieto2015}, and coincides with its emission  rate by the electric and/or magnetic dipole induced on the particle, as should be.

3. The extinction torque obtained from  the spin  angular momentum  conservation contains two terms: one is the  so-called intrinsic torque by several authors so far, usually employed in most studies on dipolar particles based on the static theory, and which however we show to be cancelled by a part of the total recoil torque, which  is also intrinsic as it does not depend on the origin of coordinates; the other term comes from the spatial structure of the incident field which, in turn, is equal to twice the corresponding extinction torque yielded by the conservation of orbital angular momentum. If the incident wave is plane, obviously there is no term due to the  field spatial structure; hence the orbital angular momentum conservation yields no extinction part, which is consistent with the absence of incident orbital angular  momentum in such a wave. Also, in this latter case, the above mentioned cancellation conveys a transfer of spin angular momentum to the particle  accounted for by the remaining  part of  the recoil torque through absorption of the incident energy. 

 4. Therefore, if the incident field is a  plane wave, according to Point 3 above the existence of a recoil torque stemming from the conservation of orbital angular momentum has to come from the incident spin angular momentum, this being a manifestation of the {\it spin-orbit interaction} which is included in the conservation laws.

5. For a circularly polarized incident plane wave, if the particle produces a circularly polarized scattered field, (for which it is sufficient that its electric and magnetic polarizabilities coincide), then the electromagnetic  torque  and force on that particle become proportional to each other.

6. Taking into account the spatial structure of the field, the torques derived from the spin and orbital conservation laws admit a decomposition into conservative and non-conservative components, analogous to that of electromagnetic forces. There is a {\it gradient component} of the torque, where now the wavefield {\it helicity} plays a  role analogous to that of the energy for the optical force; and hence may be employed as a {\it rotational optical tweezer} that aligns the  torque acting on the particle along an equilibrium direction. Likewise, there is a component given by the {\it spin angular momentum}, (which coincides with the helicity flow in the dual-symmetric formulation for quasi-monochromatic fields \cite{bliokh2} here employed), analogous to the radiation pressure, or Poynting vector part,  of the force. In addition, there are other non-conservative components  that describe torques due to circulation of field vortices around the dipolar object.

7. In contrast with plane waves,  incident fields with certain spatial structures, and some beams in particular, may easily produce negative electromagnetic torques on one single small particle.

\section{The flow of  helicity}

We shall address time-harmonic fields with electric and magnetic vectors:  ${\bf \cal E}({\bf r},t)=\Re [{\bf E}({\bf r}) \exp(-i\omega t)]$ and  ${\bf \cal B}({\bf r},t)=\Re [{\bf B}({\bf r}) \exp(-i \omega t)]$, a functional form that also applies to all potentials and  currents. $\Re$ denotes real part. Although such fields are not source-free \cite{sherman,nietolib}, they will be considered in regions with no  sources  so that they admit an angular spectrum representation of plane wave components \cite{nietolib,mandel}, either convergent, divergent, or both, satisfying at $kr \rightarrow \infty$ Sommerfeld's radiation condition. Such fields are of course transversal, namely, they hold  Maxwell's equations: $\nabla \cdot {\bf \cal E}=0$,  $\nabla \cdot {\bf \cal B}=0$.

In a non-absorbing dielectric medium of  refractive index $n=\sqrt{\epsilon \mu}$, ($\epsilon$ and $\mu$ representing the dielectric permittivity and the magnetic permeability), the helicity density $\mathscr{ H}$,  the flow of helicity density $\mathscr{ F}$, and the tensor ${\tens}_{ij}$ of density of flow of helicity  flux, of this light field are defined in the dual-symmetric formulation as (cf. \cite{bliokh2,cameron1,nori}):
\begin{equation}
\hel=\frac{1}{2}(\frac{1}{\mu}{\bf \cal A}\cdot\bf{\cal B}-\epsilon {\bf \cal C} \cdot \bf \cal E), \label{h}
\end{equation}
\begin{equation}
\flow=\frac{c}{2 \mu}({\bf \cal E}\times\bf{\cal A} +  {\bf \cal B} \times \bf \cal C), \label{f111}
\end{equation}
\be
 {\tens}_{ij}=\frac{c^2}{2\mu}[ {\cal E}_i {\cal C}_j + {\cal E}_j {\cal C}_i -\frac{1}{\epsilon \mu}( {\cal A}_i {\cal B}_j+ {\cal A}_j {\cal B}_i) + \nonumber \\
 \delta_{ij} (\frac{1}{\epsilon \mu}{\cal A} \cdot {\cal B} - { \cal E} \cdot { \cal C})], \,\,\, (i,j=1,2,3).
 \label{tens}
\ee
In (\ref{h}) and (\ref{f111})   ${\bf \cal A}$ and ${\bf \cal C}$ are vector potentials  such that ${\bf \cal B}=\nabla \times {\bf \cal A}$ and  ${\bf \cal E}=-\nabla \times {\bf \cal C}$
 and are transversal in a Coulomb's gauge \cite{cameron1}: $\nabla\cdot {\bf \cal A}=\nabla\cdot {\bf \cal C}= 0$. From Maxwell's equations: 
\be
\dot{\bf \cal A}= -c {\bf\cal  E}, \,\,\,\,\,
\dot{\bf \cal C}= - \frac{c}{\epsilon\mu} \nabla\times {\bf \cal A } + \frac{4 \pi}{\epsilon}  {\bf \cal K }.
\label{pots}
\ee

The upper dot meaning $\partial_t$, $c$ being the light speed in vacuum. In writting $\dot{\bf \cal C}$ in (\ref{pots})  we have taken into accont that   the electric current density, which we denote as $\bf \cal J$,  is transversal since the existence of ${\bf \cal A}$ and the law $\nabla\cdot \epsilon {\bf \cal E}= 4 \pi \rho$ convey that the electric charge density $\rho$ and any (static) scalar potential are zero \cite{jackson}, and thus ${\bf \cal J}$   has been expressed as $ {\bf \cal J}= \nabla \times {\bf \cal K }$.

The transversality of  $ {\bf \cal A }$ and $ {\bf \cal C }$ also amounts through   the two first Eqs. (\ref{pots}) to that of ${\bf \cal E }$ and   ${\bf \cal B }$, and hence is compatible with the above  argumentation  based on their angular spectrum  \cite{mansu}, fulfilled by most optical wavefields outside near-field regions. In particular, this also involves that any far-zone scattered  or radiated field, which is given by the propagating part of its angular spectrum  \cite{nietolib,mandel},  is gauge-invariant. This agrees with \cite{mansu,marston1} and  implies that  since  both  ${\bf \cal E }$ and  ${\bf \cal B }$ are gauge invariant, the angular momentum ${\bf J}$ of these fields admits a decomposition  ${\bf J}={\bf L}+{\bf \flow}$ into an orbital  ${\bf L}$, [proportional to $\epsilon{\bf \cal E}\cdot({\bf r}\times \nabla){\bf \cal A}+\mu^{-1}{\bf \cal B}\cdot({\bf r}\times \nabla){\bf \cal  C}$] and a spin  ${\bf \flow}$, [proportional to $\epsilon{\bf \cal E} \times{\bf \cal A}+\mu^{-1}{\bf \cal B} \times {\bf \cal  C}$, see above], angular momenta;  both $ {\bf L}$ and ${\bf \flow}$  also being  gauge-invariant.  

Therefore, it  makes sense to study the conservation laws for the spin and orbital angular momenta separately. However,  as we shall discuss in one example, sometimes the transversality condition is not taken into account, dealing with some representations of optical beams which do not fulfill it, thus preventing their inclusion in such decomposition of the angular momentum and, in general, in any theory based on such transversality equations.

 From the above conditions, the definitions, $\hel$, $\flow$ and  ${\tens}_{ij}$ hold the continuity equations \cite{bliokh2,cameron1}
\begin{equation}
\dot{\hel}+ \nabla \cdot \flow =- \perd , \label{contH}
\end{equation}
\begin{equation}
\dot{\flow}+ \nabla \cdot \tens_{ij} =- \perdr , \label{contF}
\end{equation}
Where the dissipation on transmission of spin flow by interaction of the fields with matter is given by $\perd= 2\pi ( {\bf  \cal E}\cdot {\bf  \cal K}-{\bf \cal J} \cdot {\bf \cal  C})$ and $\perdr=( 2\pi c/\epsilon \mu)( {\cal  K}\times {\cal B} + {\cal J}\times  {\cal  A})$, respectively .

Eqs. (\ref{contH}) and (\ref{contF}) are  conservation equations for the helicity and its flow, respectively, and are  as fundamental as those for the energy,  linear and angular momenta. We shall concentrate in (\ref{contF}), where  $\perdr$ is linked to the absorption by the particle, which gives rise, as shown later, to the mechanical action that it experiences as a consequence of the conservation of the helicity flow.

The time-harmonicity of the fields and potentials converts the quantities of Eqs. (\ref{pots})-(\ref{contF}) into
\be
{\bf  A}= - \frac{i}{k} {\bf E}, \,\,\,\,
{\bf  C}= - \frac{i}{\epsilon} [\frac{{\bf B}}{k \mu} - \frac{4 \pi}{\omega} {\bm K}],  \label{potsr}
\ee
and the time-averged (denoted by $<\cdot>)$:
\begin{equation}
\hel= <\hel>=\frac{1}{2k}\sqrt{\frac{\epsilon}{\mu}} \Im ({\bf E}\cdot {\bf B}^*), \label{hh1}
\end{equation}
\begin{equation}
\flow= <\flow> =  \frac{c}{4n k } \Im ( \epsilon {\bf E}^* \times {\bf  E}  + \frac{1}{\mu}{\bf  B}^*\times{ \bf B})={\flow}_{e}+{\flow}_{m}  . \label{ff1}
\end{equation}
${\flow}_{e}$  and $ {\flow}_{m}$ being the electric and magetic parts of ${\flow}$ which now coincides with the spin angular momentum  density.

And
\be
 {\tens}_{ij}= <{\tens}_{ij}>=\frac{c^2}{2kn\mu} \Im[  E_{i}^{*} B_{j}+E_{j}^{*} B_{i} -\delta_{ij} {\bf E}^{*} \cdot {\bf B}], \,\,\, \label{tensrnn}
\ee
which is the density of spin angular momentum flow tensor.  $ i,j=1,2,3$, $\Im$  denotes imaginary part and $k=n \omega/c$;  and  we have written  ${\bf \cal A}= \Re[ {\bf A}({\bf r}) \exp (-i \omega t)]$, 
${\bf \cal C}= \Re[{\bf C}({\bf r})  \exp (-i \omega t)]$,   ${\bf \cal J}= \Re[{\bm J}({\bf r})  \exp(-i \omega t)]$,  ${\bf \cal K}=\Re[ {\bm K}({\bf r})  \exp(-i \omega t)]$. Also $\perd$ in (\ref{contH})  and $\perdr$ in (\ref{contF}) are replaced by the time-averaged quantities:
\begin{eqnarray}
< \perd >=\pi [\frac{2}{ck}\sqrt{\frac{\mu}{\epsilon}} \nabla\cdot \Im ({\bm K}\times {\bf B}^*)  \nonumber \\ -\frac{1}{kn} \Im ( {\bm J} \cdot {\bf  B}^*)  +\frac{4\pi}{ck}\sqrt{\frac{\mu}{\epsilon}}\Im  ({\bm J}\cdot {\bm K}^*)  
 +\Re  ({\bf E}\cdot {\bm K}^*) ]  . \label{pp}
\end{eqnarray}
\begin{eqnarray}
< \perdr_i >=-\frac{\pi c}{\epsilon \mu} \{ c \Im ({\bf E}\times {\bm J}^*)_{i}+\Re ({\bf B}\times {\bm K}^* )_{i}+\nonumber \\
\frac{2\mu}{c}\partial_{j}[\delta_{ij} \Im ( {\bf E} \cdot {\bm  K}^*) -\Im( E_{i} K_{j}^{*}+E_{j} K_{i}^{*})]\} . \label{rr}
\end{eqnarray}
At this point it is pertinent to comment that in the case dealt here of time-harmonic fields, Maxwell's  equations and the above relations show that (\ref{hh1}) and (\ref{ff1}) are proportional to Lipkin's zilches \cite{lipkin}, used in recent works as the field chirality $\qui$  and flow of chirality $\s$ \cite{tang1,tang2}, as well as  the  tensor of spin flow  ${\tensq}_{ij}$, \cite{cameron1}:
\begin{equation}
\qui= <\qui>=k^2 \hel= k^2 <\hel>\label{kk}
\end{equation}
\begin{equation}
\s= <\s> = k^2 \flow= k^2 <\flow>\label{ss}
\end{equation}
\begin{equation}
\tensq_{ij}=<\tensq_{ij}>=k^2 \tens_{ij}=k^2  <\tens_{ij}> .  \label{nn}
\end{equation}
The dissipative terms that appear in the continuity equations for  $\qui$,  $\s$ and  ${\tensq}_{ij}$, are however different from those of  (\ref{contH}) and   (\ref{contF}).
Ref. \cite{cameron1} argues that  $\flow$ and $\tens_{ij}$ are the quantities with dimensions of angular momentum, rather than the zilches; and $<\hel>$, $<\flow>$ and $<\tens_{ij}>$ are those   to deal with in mesurements. We follow this, although for these monochromatic fields both kind of magnitudes should, as shown in (\ref{kk}) - (\ref{nn}), lead to the same interpretation of effects.

\section{Scattering and the conservation of  spin angular momentum}
We integrate (\ref{contF}) in any volume $v$ that contains the object with its distribution of currents. The divergence term is transformed into the flow of an entity across the  surface $\Sigma$ of  this volume that will eventually be taken as a large sphere. 

Let us first consider a  monochromatic, elliptically polarized, plane wave incident on a body of arbitrary form enclosed in $v$. (cf. Fig. 1). The field at any point of the surrounding medium may be represented as the sum of the incident  and scattered fields, thus its space dependent vectors are:  ${\bf E}({\bf r})={\bf E}^{(i)}({\bf r})+{\bf E}^{(s)}({\bf r})$,   ${\bf B}({\bf r})={\bf B}^{(i)}({\bf r})+ {\bf B}^{(s)}({\bf r})$.

\begin{figure}[htbp]
\centerline{\includegraphics[width=1.0\columnwidth]{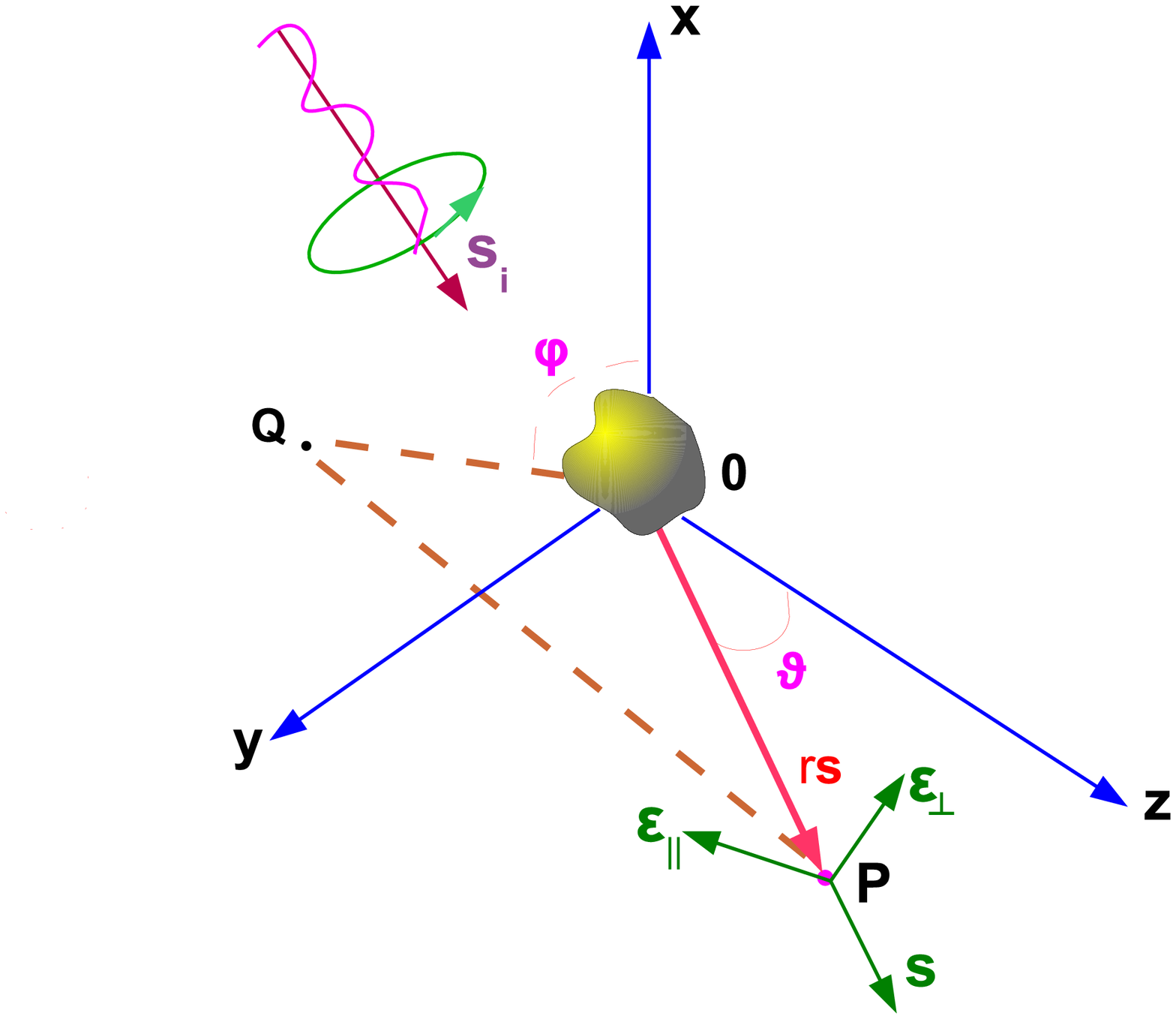}}
\caption{(Color online). Illustration of an elliptically polarized plane wave incident on a  polarizable particle of arbitrary shape. The fields are evaluated at the point {\bf P}, of  coordinates   $(r, \theta, \phi)$ and  position vector ${\bf r}=r \bf s$, (${\bf s}$ being the unit vector along ${\bf r}$),  which eventually belongs to a sphere of integration   of radius $r$, centered at some point  ${\bf r}_ {0}$  of the particle.  ${\bf r}_ {0}$ acts as the framework center {\bf 0}. The point {\bf Q} is the projection of {\bf P} on the plane $ OXY$; the scattering plane being $OPQ$.  We  show the three orthonormal vectors:  ${\bf s}$,  ${\bm \epsilon}_{\parallel}$ (in the plane $OPQ$ and in the sense of rotation of $\theta$), and ${\bm \epsilon}_{\perp}$ (normal to $OPQ$).}
\end{figure}
The incident field is written as  ${\bf E}^{(i)}={\bf e}^{(i)} e^{ik({\bf s}_{i}\cdot {\bf r})}$,  ${\bf B}^{(i)}={\bf b}^{(i)} e^{ik({\bf s}_{i}\cdot {\bf r})}$,   and that scattered in the far zone:  ${\bf E}^{(s)}={\bf e}({\bf s}) \exp(ikr)/r$, ${\bf B}^{(s)}={\bf b}({\bf s}) \exp(ikr)/r$ . So that of course:  ${\bf b}^{i}=n {\bf s}_{i} \times  {\bf e}^{i}$, ${\bf e}^{i} \cdot {\bf s}_{i}= {\bf b}^{i} \cdot {\bf s}_{i}=0$;  ${\bf b}=n {\bf s} \times  {\bf e}$, ${\bf e} \cdot {\bf s}= {\bf b} \cdot {\bf s}=0$.

 Then the   time-averaged  density of spin angular momentum flow may be written: ${\tens}_{ij}= {\tens}_{ij}^{(i)}+ {\tens}_{ij}^{(s) }+{\tens}_{ij}' $ .  Where
\begin{eqnarray}
{\tens}_{ij}^{(i)}=\frac{c^2}{2kn\mu} \Im[  E_{i}^{(i) *} B_{j}^{(i)}+E_{j}^{(i) *} B_{i}^{(i)} \nonumber \\
 -\delta_{ij} {\bf E}^ {(i) *} \cdot {\bf B}^{(i)}]  .  \,\,\,\,\,\,\,\,\,\,\,\,   (10a)  \nonumber\\
{\tens}_{ij}^{(s)}=\frac{c^2}{2kn\mu} \Im[ E_{i}^{(s) *} B_{j}^{(s)}+E_{j}^{(s)*} B_{i}^{(s)} \nonumber \\
 -\delta_{ij} {\bf E}^{(s)  *} \cdot {\bf B}^{(s)} ] .  \,\,\,\,\,\,\,\,\,\,\,\,\nonumber (10b) \\ 
{\tens}_{ij}^{'}=\frac{c^2}{2kn\mu} \Im[E_{i}^{(i) *} B_{j}^{(s)}+E_{i}^{(s) *} B_{j}^{(i)} \nonumber \\
+E_{j}^{(i) *} B_{i}^{(s)}+E_{j}^{(s) *} B_{i}^{(i)} \nonumber \\
 -\delta_{ij}( {\bf E}^{(i)  *} \cdot {\bf B}^{(s)}  +{\bf E}^{(s)  *} \cdot {\bf B}^{(i)} ] .\,\,\,\,\,\,\,\,\,\,\,\,(10c) \nonumber
\end{eqnarray}

From Eq.(\ref{contF}) we write  the rate ${\cal L}$ at which this density flows   on interaction of the incident wave with the body, in terms of its outward flux:   ${\cal L}=-(n^2/ 4\pi c^2) \int_{\Sigma}  d\Omega r^2 {\tens}_{ij} \cdot {\bf s}$   through the surface  $\Sigma$ of any large sphere of radius $r$ with center at some point ${\bf r}_0$ of the body. $d\Omega$ is the element of solid angle  and ${\bf s}$ denotes the outward normal.  I.e., according to Eq.(\ref{contF}):
\begin{equation}
 {\cal L}=  {\cal L}^{(i)}+ {\cal L}^{(s)}+{\cal L}'. \label{Lho1}
\end{equation}
${\cal L}^{(i)}$, ${\cal L}^{(s)}$ and ${\cal L}'$ are respectively the $\Sigma$  integrals of the projections on $\bf s$ of  $ -{\tens}_{ij} ^{(i)}$, $- {\tens}_{ij} ^{(s)}$ and $- {\tens}_{ij} ' $ on the surface of the sphere, respectively.  Of course from Eq.(\ref{contF}): ${\cal L}=(n^2/ 4\pi c^2)\int_{\Sigma} d r d\Omega r^2 <\perdr >$.

From these equations   $ {\cal L}^{(i)}=0$, so that  (\ref{Lho1}) becomes
\be
 {\cal L}= {\cal L}^{s}+{\cal L}'. \label{lh1}
\ee
The integrals  of $-{\tens}_{ij}^{s} \cdot {\bf s}$ and $-{\tens}_{ij}' \cdot {\bf s}$ across $\Sigma$  leave the following non-zero terms, [cf. (\ref{tensrnn})]:
\be
{\cal L}^{(s)}=- \frac{n^2}{ 4 \pi c^2}\int_{\Sigma}  d\Omega r^2  {\tens}_{ij}^{(s)} s_j \nonumber \\
= \frac{n}{8 \pi k \mu}\int_{\Sigma}  d\Omega r^2  \Im[ {\bf E}^{(s)  *} \cdot {\bf B}^{(s)} ] {\bf s}=\,\, \,\,\,\,\,\,\, \nonumber \\
-\frac{n^2}{8 \pi k\mu}\int_{\Sigma}  d\Omega  \Im \{ {\bf s}\cdot [  {\bf e}^{(s) *} ({\bf s}) \times  {\bf e}^{(s)}({\bf s})] \}{\bf s}= \,\,\,\,\,\,\,\,\,\nonumber \\ 
-\frac{1}{8 \pi k\mu}\int_{\Sigma}  d\Omega  \Im \{ {\bf s}\cdot [  {\bf b}^{(s)  *}({\bf s}) \times  {\bf b}^{(s)}({\bf s})] \}{\bf s}= \,\,\,\,\,\,\,  \nonumber \\
-\frac{1}{16 \pi k}\int_{\Sigma}  d\Omega   
\Im \{\epsilon {\bf s}\cdot [  {\bf e}^{(s)  *}({\bf s}) \times  {\bf e}^{(s)}({\bf s})] \,\,\,\,\,\,\,\,\,\,\,   \nonumber \\ 
+\frac{1}{\mu} {\bf s}\cdot [ {\bf b}^{(s ) *}({\bf s}) \times  {\bf b}^{(s)}({\bf s})] \} {\bf s}. \,\,\,\,\,\,\, \,\,\,\,\,\,\,  \label{tss1}
\ee
and
\be
{\cal L}'=- \frac{n^2}{ 4 \pi c^2} \int_{\Sigma}  d\Omega r^2    {\tens}_{ij}'    s_{j}=\nonumber \\
\frac{n}{8 \pi k \mu}\int_{\Sigma}  d\Omega R^2  \Im[ {\bf E}^{(i ) *} \cdot {\bf B}^{(s)}  +{\bf E}^{(s ) *} \cdot {\bf B}^{(i)}]{\bf s} =\nonumber \\
-\frac{n}{2 \mu  k^2 } \Re[{\bf B}^{(i)  *}({\bf r}_0)  \cdot  {\bf e}({\bf s}_{i})] {\bf s}_{i}= \nonumber \\
\frac{n }{2 \mu k^2 } \Re[{\bf E}^{(i) *}({\bf r}_0)  \cdot  {\bf b}({\bf s}_{i})]{\bf s}_{i} . \label{fit1}
\ee
In deriving (\ref{fit1}) we have used Jones' lemma based on the principle of the stationary phase \cite{jones,born}:
\be
\frac{1}{r}\int  d\Omega r^2  F({\bf s}) e^{-ik ({\bf s}_{i} \cdot {\bf s})r} \sim \frac{2\pi i}{k}[F({\bf s}_{i})e^{-ik r} \nonumber \\
 - F(-{\bf s}_{i})e^{ik r} ] . \label{psp1}
\ee
 Notice from (\ref{ff1}) and (\ref{tensrnn}) that all these quantities of the conservation law  (\ref{lh1}) -  (\ref{fit1}) are also time-averages.  Eq. (\ref{tss1}) conveys the ellipticity  of the electric or the magnetic field, as well as the dual-symmetric one containing both ${\bf e}$ and ${\bf b}$.  This electric-magnetic duality is also seen in (\ref{fit1}). 

 Eq.(\ref{lh1}), together with (\ref{tss1})  and (\ref{fit1}),  constitutes the law for the moment of the force  exerted by the incident wave on the object, expressed in terms of the  flux ${\cal L}$ of the spin  flow tensor ${\tens}_{ij}$ across any closed surface surrounding it. As seen, this stems from the conservation of the spin angular momentum density and, like the optical force \cite{MNV2010} it has a  contribution ${\cal L}^{(s)}$  of the scattered field plus one ${\cal L}'$  arising from the  interference between this scattered field and the incident one. 

This force moment ${\cal L}$ constitutes the  contribution to the {\it electromagnetic torque} on the scatterering object (acting the ellipticity of the scattered field as a factor in the integration) by  extinction of the incident spin angular momentum, and its transfer to the body at a rate proportional to a certain amplitude component  of the  scattered wave, projected along the  polarization   of the incident wave  as a result of their mutual interference in the forward direction ${\bf s} = {\bf s}_i$.

\section{Consequences for a magnetodielectric bi-isotropic dipolar particle}
\subsection{Incident plane wave}
Let us consider a magnetodielectric bi-isotropic  particle, dipolar in the wide sense, i.e. characterized by its polarizabilities: electric $\alpha_{e}$ ,  magnetic  $\alpha_{m}$, and magnetoelectric $\alpha_{em}$,  $\alpha_{me}$, such that for example if  it is a sphere  these quantities are given by the first order Mie coefficients $a_{1}$, $b_{1}$ and  $c_{1} $ as
$\alpha_{e}=i\frac{3}{2k^{3}}a_{1}$,  $\alpha_{m}=i\frac{3}{2k^{3}}b_{1}$,  $\alpha_{em}=i\frac{3}{2k^{3}}c_{1}$,  $\alpha_{me}=i\frac{3}{2k^{3}}d_{1}$, respectively \cite{MNV2010,Nieto2011,chanlat}. Then the electric and magnetic dipole moments, $ {\bf p}$ and  ${\bf m}$, on the particle, induced by the incident field are:
\be
{\bf p}=\alpha_{e} {\bf E}^{(i)}+\alpha_{em}{\bf  B}^{(i)}, \,\,\,\,\,
{\bf m}=\alpha_{me}{\bf E}^{(i)}+\alpha_{m}{\bf B}^{(i)}. \label{consti1}
\ee
If, in addition,  such sphere is chiral, then    $c_{1}=-d_{1}$  \cite{chanlat} and hence $\alpha_{em}=-\alpha_{me}$. 

The amplitudes of  the fields  scattered by this particle in the far-zone are:
\begin{equation}
{\bf e}({\bf s})=k^2 \frac{e^{ikr}}{r} [\epsilon^{-1}({\bf s} \times {\bf
p})\times {\bf s}- \sqrt{\frac{\mu}{\epsilon}}({\bf s}\times
{\bf m})], \label{dipe1}
\end{equation}
\begin{equation}
{\bf b}({\bf s})=k^2 \frac{e^{ikr}}{r}[\mu({\bf s} \times {\bf m})\times
{\bf s}+\sqrt{\frac{\mu}{\epsilon}} ({\bf s}\times {\bf p})]. \label{dipm1}
\end{equation}
Introducing (\ref{dipe1}) and (\ref{dipm1}) into  (\ref{tss1}) and (\ref{fit1}), evaluating the angular integrals, and substituting the results in (\ref{lh1}), we obtain:
\be
{\cal L}^{(s)}=-\frac{k^3}{6} \Im \{\frac{1}{\epsilon} {\bf p}^{*}  \times {\bf p} + \mu {\bf m}^{*}  \times {\bf m} \}. \label{ttss1}
\ee
and
\be
{\cal L}'=-
\frac{n}{2 \mu  }
 \Re [(\frac{1}{\epsilon}   {\bf p} \cdot {\bf B}^{(i) *}   - \mu    {\bf m} \cdot {\bf E}^{(i) *} 
){\bf s}_{i}] .  \label{fiitt1}
\ee
Eq.(\ref{ttss1}) {\it is half the rate of radiated angular momentum} by the dipolar object \cite{jackson}, or equivalently half  the {\it  recoil optical torque} on the particle which one would obtain from the conservation of  the total angular angular  momentum   by integration of  Maxwell's stress tensor $T_{ij}$: $\int_{\Sigma} dR d\Omega  r{\bf s} \times T_{ij} s_j$  \cite{nieto2015}.

Interestingly, since ${\bf p}\cdot {\bf s}_{i}={\bf m}\cdot {\bf s}_{i}=0$, (\ref {fiitt1}) may also be expressed as:
\be
{\cal L}'=<{\bm \Gamma}_{0}>.  \label{torq11}
\ee
 $<{\bm \Gamma}_{0}>$ is the {\it extinction electromagnetic torque}, analogous to the extinction energy in the optical theorem \cite{MNV2010, chen1,born},
\be
<{\bm \Gamma}_{0}>=
\frac{1}{2  } \Re [ {\bf p}\times {\bf E}^{(i )*}  +  {\bf m}\times{\bf B}^{(i) *} ] .  \label{torq21}
\ee
 Sometimes  $<{\bm \Gamma}_{0}>$ is called "the intrinsic torque"  \cite{chaumet, cana}  exerted by the incident plane wave on the particle. However, ${\cal L}^{(s)}$, Eq.(\ref{ttss1}), is seen to be also an intrinsic torque component.

It is useful to consider a Cartesian framework (cf. Fig.1)  where  the elliptically polarized  incident plane wave has ${\bf s}_i=(0,0,1)$. Its electric vector being written in an helicity basis ${\bm \epsilon}^{\pm}=(1/\sqrt{2})(1, \pm i, 0)$ as the sum of a left-hand   and a right-hand  circularly polarized plane wave, so that ${\bf e}_{i}=( e_{i x},e_{i y},0)=e_{i}^{+}{\bm \epsilon}^{+}+ e_{i}^{-}{\bm \epsilon}^{-}$ and  ${\bf  b}_{i}= ( b_{i x},b_{i y},0)=n ( -e_{i y},e_{i x},0)=b_{i}^{+}{\bm \epsilon}^{+}+ b_{i}^{-}{\bm \epsilon}^{-}=-ni(e_{i}^{+}{\bm \epsilon}^{+}- e_{i}^{-}{\bm \epsilon}^{-})$. The sign $\pm$ standing for LCP (+) and RCP (-), respectively. In such representation, the incident helicity acquires the form:    $ {\hel^{i}}=(\epsilon/k)\Im[e_{i x}^{*}e_{i y}]= (\epsilon/2k) S_{3}=(\epsilon/2k)[|e_{i}^{+}|^2-|e_{i}^{-}|^2]$, which clearly expresses this magnitude as the difference between left circular and right circular intensities of the field, extensively exploited in dichroism \cite{tang1,tang2,schellmann}  and enantiomeric molecule discrimination.  $S_3 = 2 \Im[e_{i x}^{*}e_{i y}]= |e_{i}^{+}|^2-|e_{i}^{-}|^2$ represents the fourth Stokes parameter \cite{marston1,born}.

In the case of circular polarization of the incident wave,   $e_{ix}=e$, $e$ is real, and  $e_{iy}=\pm i e$ depending on whether the light is left  or right circular; also $n{\bf e}_{i} = \pm i {\bf b}_{i} $; the incident helicity is $ {\hel^{i}}=\pm e^2/k$. In addition, when the particle satisifies the first Kerker condition \cite{kerk,suppre,greffin}, thus yielding zero differential scattering cross section in the backscattering direction,  $\alpha_e /\epsilon= \mu \alpha_m$, then ${\bf p}=\pm i n {\bf m}$ and    ${\bf b}({\bf s})=\mp  i  n {\bf e}({\bf s})$; 
 i.e. the scattered field is circularly polarized  with respect to the Cartesian system of orthonormal vectors: $({\bm \epsilon}_{\perp},{\bm \epsilon}_{\parallel}, {\bf s})$, (see Fig.1),   ${\bm \epsilon}_{\perp} $ and  ${\bm \epsilon}_{\parallel}$ being, respectively,  perpendicular and parallel to the scattering plane \cite{nietoPRA1}. 

In consequence, in left circular polarization one has that  ${\bf e}({\bf s})=({\bf e}({\bf s}) \cdot {\bm \epsilon}_{\perp})(1,\pm i, 0)$ and ${\bf b}({\bf s})=( n {\bf e}({\bf s}) \cdot {\bm \epsilon}_{\perp})  (\mp i, 1,0)$.  The helicity of the scattered field being proportional to its intensity: ${\hel^{s}}=\pm \frac{\epsilon}{2k} |{\bf e}({\bf s})|^2$, so that since then the flow of helicity becomes proportional to that of energy, one has that  the torque ${\cal L}$ , Eq.(\ref{lh1}), and the flow tensor  $ {\tens}_{ij}$,Eq. (\ref{tensrnn}), become proportional to the time-averaged force $<{\bf F}>$ and Maxwell's stress tensor $ <T_{ij}>$, \cite{jackson, MNV_PTRSL}:  $ {\cal L}= \pm (1/ k) <{\bf F}>$,  $- {\tens}_{ij} =\pm (1/ k)  <T_{ij}>$. This matches with a statement of \cite{cameron1} on the existence of a mapping of the helicity into the energy for circularly polarized waves.
\subsection{Arbitrary incident wave}

The above equations are generalized for arbitrary illuminating wavefields,  expressible in terms of its angular spectrum  of plane wave components \cite{nietolib,mandel}: 
\be
{\bf E}^{(i)}({\bf r})=\int_{\cal D} {\bf e}^{(i)}({\bf s}) e^{ik({\bf s}\cdot {\bf r})} d\Omega, \nonumber \\
{\bf B}^{(i)}({\bf r})=\int_{\cal D} {\bf b}^{(i)}({\bf s}) e^{ik({\bf s}\cdot {\bf r})} d\Omega.  \label{ang111}
\ee
The integration being done in the unit sphere ${\cal D}$ whose element of solid angle is $d \Omega$.  In this regard, notice that  Eqs.(\ref{dipe1}) and (\ref{dipm1}) now  constitute the angular spectra of the electric and magnetic scattered vectors when they are represented  in the form (\ref {ang111}).  This integration contains both propagating and evanescent waves \cite{nietolib,mandel,MNV2010};  and to include them  both, ${\bf s}_i$ in  (\ref{fiitt1}) must be replaced by  ${\bf s}_{i}^{*}$, complex conjugated  of  ${\bf s}_{i}=(s_{i}^{x},
s_{i}^{y}, s_{i}^{z}) $, where  $s_{i}^{z}= \sqrt{ 1-(s_{i}^{x 2}+s_{i}^{y 2})}$  if  $ s_{i}^{x 2}+s_{i}^{y 2} \leq 1$, (propagating components); and   $s_{i}^{z}= i \sqrt{ (s_{i}^{x 2}+s_{i}^{y 2})-1}$ if   $s_{i}^{x 2}+s_{i}^{y 2} > 1$, (evanescent  components) .

From (\ref{ang111}), using the same procedure as before for each plane wave component, and summing up all of them,  an easy calculation shows that  ${\cal L}^{s}$ remains as in 
 (\ref{ttss1}), however the extinction torque ${\cal L}'$ now becomes
\be
{\cal L}'= 
\frac{n}{2 k \mu} \Im \{\frac{1}{\epsilon} p_j \partial_i B_{j}^{(i )*}- \mu m_j \partial_i E_{j}^{(i) *} \} ;  \label{lprimfin1}
\ee
which  may be also expressed as
\be
 {\cal L}'=< {\bm  \Gamma}_0>
+\frac{1}{2 k} \sqrt{\frac{\epsilon}{ \mu}}\Im\{\frac{1}{\epsilon}({\bf p} \cdot\nabla){\bf B}^{(i) *}-\mu ({\bf m} \cdot\nabla){\bf E}^{(i) *}\} .\,\,  \label{torq31}
\ee
Where $<{\bm \Gamma}_0>$  now  is
\be
<{\bm \Gamma}_0>=
\frac{1}{2  } \Re [ {\bf p}\times {\bf E}^{(i) *} +  {\bf m}\times{\bf B}^{(i) *} ] .  \label{torq2111}
\ee
The fields ${\bf E}^{(i)}$  and ${\bf B}^{(i)}$ being given by  Eqs. (\ref{ang111}).

At this stage it is worth observing how taking (\ref{consti1}) into account, Eq.(\ref{lprimfin1}) confers, after separating real and imaginary  parts, a physical meaning in terms of the extinction torque ${\cal L}'$ to some quantities associated to 'magnetoelectric  effects' introduced in  \cite{bliokhmagnetoel}. 

\subsection{On the  torque involved in  the conservation of the  spin angular momentum}
The above equations  not only show that the ${\cal L} $ quantities  have torque dimensions, as they should, but as mentioned before they   manifest that  ${\cal L}^s $ is also an intrinsic torque. Notice on comparing (\ref{torq31}) and (\ref{torq2111})   with (\ref{torq11}) and (\ref{torq21}), the two additional terms in (\ref{torq31}) due to the spatial structure of the incident field when this is not a plane wave. This term accounts for an extrinsic torque component. Also, the second and third terms of  (\ref{torq31}) have an analogy with those of the dipolar component of the electromagnetic force, whereas those contained in $<{\bm \Gamma}_0>$ keep it with those of the Lorentz's component of the optical force. This duality between the ${\bf E}$ and ${\bf B}$ vectors is also evident by comparing   (\ref{lprimfin1}) with the expression of the  electric ($e$) plus magnetic ($m$) forces on a magnetodielectric dipolar particle \cite{MNV2010}.  In this connection, the {\it recoil} optical  torque, Eq.(\ref{ttss1}),  due to interference of the scattered fields and which, as said, is  also valid for an arbitrary incident field, has a formal analogy with the electric-magnetic dipole interaction electromagnetic  force:  $ {\bf F}_{e - m}$ \cite{MNV2010}.

In summary, the  torque   on the particle ${\cal L}=  {\cal L}^s +  {\cal L}'$, Eq.(\ref {lh1}), governed by the conservation of the spin angular momentum, is given by  the sum of   (\ref{ttss1}) plus (\ref{lprimfin1}) or (\ref{torq31}). This torque has the  recoil component  ${\cal L}^s$  added to ${\cal L}'$, which  is as necessary to describe the dynamics as the above referenced $ {\bf F}_{e - m}$ component of the force .

 However this scattering torque (\ref{ttss1})  is only half the recoil torque exerted by the field on the particle, obtained from the conservation of total angular momentum \cite{nieto2015}. As we shall see, the other half is obtained from the conservation  of the orbital angular momentum. As expected from the premises above adopted in this formuation, this  is consistent with the addition of both angular momenta to describe the total  transfer to the body.

\section{Decomposition of the  torque  involved in  the conservation of spin angular  momentum into conservative and non-conservative components}
Like the electromagnetic force \cite{MNV2010,albala,berry}, the optical torque is amenable of a  decomposition into conservative and non-conservative components, as shown next.

At this point we emphasize again that  the vectors ${\bf E}^{(i)}$ and ${\bf B}^{(i)}$ satisfy  the transversality condition $\nabla\cdot{\bf E}^{(i)}=0$ and $\nabla\cdot{\bf B^{(i)}}=0$. From now on we omit the $(i)$ superindex. Taking in (\ref{consti1}) the real  and imaginary parts of the  polarizabilities, and  using the vector identities:
\be
\Re[({\bf B}^{*}\cdot \nabla){\bf E}]=\frac{1}{2}(Y+X); \,\,\,\Re[({\bf E}\cdot \nabla){\bf B}^{*}]=\frac{1}{2}(Y-X).  \nonumber \\
\Im[({\bf B}^{*}\cdot \nabla){\bf E}]=\frac{1}{2}(S+R);\,\,\,\,\,\,\Im[({\bf E}\cdot \nabla){\bf B}^{*}]=\frac{1}{2}(S-R). \nonumber \\
X=\Re[\nabla \times  ({\bf E} \times {\bf B}^{*})];  \,\,\,\,\,\,  \nonumber\\
Y= \Re[\nabla  ({\bf E} \cdot {\bf B}^{*})]  -kn\Im [{\bf E}^{*} \times {\bf E}]+k_0\Im [{\bf B}^{*} \times {\bf B}]; \nonumber \\
R=\Im[\nabla \times  ({\bf E} \times {\bf B}^{*})]; \,\,\,\, S= \Im[\nabla  ({\bf B}^{*} \cdot {\bf E})]; \,\, k_0 = \omega/c, \,\,\,\,\,\, \label{vectident}
\ee
  we obtain the following expressions for the  parts of  ${\cal L}={\cal L}'+{\cal L}^{s}$ ,  Eq.(\ref{lh1}). First, $ {\cal L}'$, Eqs. (\ref{lprimfin1}) or (\ref{torq31}) ,  becomes: 
\be
 {\cal L}' = \frac{n}{2 k  \mu} \{\alpha_{e}^{R} [k_0\frac{\mu}{\epsilon}\nabla{\hel} -
\frac{1}{2\epsilon} \nabla \times \Im({\bf E} \times{\bf B}^{*})]\nonumber \\
+\alpha_{m}^{R} [k_0\mu^2\nabla{\hel} +
\frac{\mu}{2} \nabla \times \Im({\bf E} \times{\bf B}^{*})]\nonumber \\
+\frac{\alpha_{e}^{I}}{2\epsilon}[\frac{4\mu k^2}{c}{\flow} 
-  \frac{8\pi \mu}{c}\nabla \times <{\bf S}>+ \nabla  \Re({\bf E} \cdot{\bf B}^{*})] \nonumber \\
+\frac{\mu \alpha_{m}^{I}}{2} [\frac{4\mu k^2}{c}{\flow}
-  \frac{8\pi \mu}{c}\nabla \times <{\bf S}>- \nabla  \Re({\bf E} \cdot{\bf B}^{*})] \nonumber \\
-8\pi\frac{\mu}{\epsilon}( \alpha_{me}^{I} \nabla <w_e>- \alpha_{em}^{I} \nabla  <w_m> )\nonumber \\
-\frac{2nk}{c}\frac{\mu}{\epsilon} ( \alpha_{me}^{R}\nabla \times  {\flow}_{e} -\alpha_{em}^{R}\nabla \times  {\flow}_{m} ) \nonumber \\
+ \frac{8\pi k \mu}{c} \sqrt{\frac {\mu}{\epsilon}}(\alpha_{me}^{R}-\alpha_{em}^{R})<{\bf S}> \}.  \label{tordip21}
\ee
The helicity ${\hel}$ and the spin angular momentum density  that appear in (\ref{tordip21}) where defined in Eqs.(\ref{hh1}) and  (\ref{ff1}), respectively. Moreover, $<w_e>=(\epsilon /16 \pi) |{\bf E}|^2$ and  $<w_m>=(1/16 \pi \mu) |{\bf B}|^2$  are the time-averaged electric and magnetic energy densities,  and  $<{\bf S}>=(c/8\pi \mu)\Re ({\bf E} \times{\bf B}^{*})$ represents the time-averaged Poynting vector. The $R$ and $I$ superindices of the polarizabilities denote their real and imaginary parts, respectively.

Notice that there are conservative terms containing $\alpha_{e}^{R} k_0 \nabla{\hel}$  and $\alpha_{m}^{R} k_0 \nabla{\hel}$, which represent a {\it  gradient optical  torque} where now the  helicity ${\hel}$ plays a role analogous to that of the field energy for the optical force.  This may constitute the basis of what we may call as a {\it rotational optical tweezer},  positioning the particle in  equilibrium  points where there is no lateral torque. This is addressed in some of the examples below. Also the imaginary parts $\alpha_{me}^{I}$ and $\alpha_{em}^{I}$ yield gradient terms of the energies, like in the gradient force components. These latter terms of the cross-polarizabilities change sign if the chirality of the particle, characterized by $ \alpha_{em}$ and $\alpha_{me}$, is reversed, a fact that suggests its use for enantiomeric separation \cite{cana}.  Likewise, there are conservative terms $\nabla  \Re({\bf E} \cdot{\bf B}^{*})$, this time with the imaginary parts  $\alpha_{e}^{I}$ and  $\alpha_{m}^{I}$.

The polarizabilities  $\alpha_{e}^{R} $  and $\alpha_{m}^{R}$ also  appear in curl  non-coservative  parts that contain the vortex   circulation of the reactive alternating flow of energy $\nabla \times \Im({\bf E} \times{\bf B}^{*})$ \cite{jackson}. These latter terms are sometimes zero, like for a highly collimated circularly polarized Gaussian beam. On the other hand, the $\alpha_{e}^{I}$ and $\alpha_{m}^{I}$ non-conservative terms proportional to ${\flow}$, come from  $<{\bm \Gamma}_0>$ in Eq.(\ref{torq31}). They  are the analogous to  the {\it  radiation pressure} terms containing the field linear momentum  in the time-averaged force. It is remarkable the analogy according to which  while in this  latter case the conservation of the wave  linear momentum leads to the radiation pressure part of the optical force, in  the optical torque the conservation of the spin angular momentum  ${\flow}$ leads to  terms that similarly contain this quantity and play the role of a {\it spinning  torque}.

There are also energy flows in the cross-polarizabilities $\alpha_{me}^{R}$ and $\alpha_{em}^{R}$  that contain  the curl of both the Poynting vector and the flow of spin angular momentum, and thus represent their respective vortices  surrounding the particle.

 Likewise, the recoil torque, Eq.(\ref{ttss1}), is written as
\be
 {\cal L}^{s} = -\frac{k^3}{6}
 \{\frac{4nk}{c}[\frac{1}{\epsilon^2}|\alpha_{e}|^{2} {\flow}_{e}+\mu^2 |\alpha_{m}|^{2} {\flow}_{m}\nonumber \\
+\frac{1}{\epsilon^2}|\alpha_{em}|^{2} {\flow}_{m}+\mu^2 |\alpha_{me}|^{2} {\flow}_{e}] \nonumber\\
+\frac{16 \pi \mu}{c}[\frac{1}{\epsilon}\Im(\alpha_{e}^{*} \alpha_{em})- \mu\Im(\alpha_{m}^{*}\alpha_{me})]<{\bf S}>\nonumber \\
-2[\frac{1}{\epsilon}\Re(\alpha_{e}^{*} \alpha_{em})+ \mu\Re(\alpha_{m}^{*}\alpha_{me})]\Im({\bf E} \times{\bf B}^{*})
 \} .  \label{tscatt11}
\ee
which in addition to non-conservative terms with electric and magnetic spins, and with the energy flux,   has conservative terms,  if the particle is bi-isotropic, containing the “reactive” or “stored energy” and its alternating flow \cite{jackson}: $\Im({\bf E} \times{\bf B}^{*})$, which  taking into account the identity:  $k\Im \{{\bf E} \times{\bf B}^{*}\}= (1/2)\nabla |{\bf E}|^2- \Re \{({\bf E}^{*} \cdot \nabla) {\bf E}$, yields gradient force-like terms, proportional to $\nabla|{\bf E}|^2$, which change sign if the particle chirality varies from left-handed to right-handed. 

Thus in summary we see that the helicity and the spin play a role in the optical torque somewhat analogous to the energy and the Poynting vector in the optical force.

In the particular case in which the particle is chiral, then \cite{tang1} $\alpha_{em}= -\alpha_{me}$ and  (\ref{tordip21})-(\ref{tscatt11}) reduce to
\be
 {\cal L}' = \frac{n}{2 k  \mu} \{\alpha_{e}^{R} [k_0\frac{\mu}{\epsilon}\nabla{\hel} -
\frac{1}{2\epsilon} \nabla \times \Im({\bf E} \times{\bf B}^{*})]\nonumber \\
+\alpha_{m}^{R} [k_0\mu^2\nabla{\hel} +
\frac{\mu}{2} \nabla \times \Im({\bf E} \times{\bf B}^{*})]\nonumber \\
+\frac{\alpha_{e}^{I}}{2\epsilon}[\frac{4\mu k^2}{c}{\flow} 
-  \frac{8\pi \mu}{c}\nabla \times <{\bf S}>+ \nabla  \Re({\bf E} \cdot{\bf B}^{*})] \nonumber \\
+\frac{\mu \alpha_{m}^{I}}{2} [\frac{4\mu k^2}{c}{\flow}
-  \frac{8\pi \mu}{c}\nabla \times <{\bf S}>- \nabla  \Re({\bf E} \cdot{\bf B}^{*})] \nonumber \\
-8\pi\frac{\mu}{\epsilon} \alpha_{me}^{I} \nabla <w> 
-\frac{2nk}{c}\frac{\mu}{\epsilon}  \alpha_{me}^{R}\nabla \times  {\flow}  \nonumber \\
+  \frac{16\pi k \mu}{c} \sqrt{\frac {\mu}{\epsilon}}\alpha_{me}^{R}<{\bf S}> \}.  \label{tordip311}
\ee
\be
 {\cal L}^{s} = -\frac{k^3}{6}
 \{\frac{4nk}{c}[\frac{1}{\epsilon^2}|\alpha_{e}|^{2} {\flow}_{e}+\mu^2 |\alpha_{m}|^{2} {\flow}_{m}\nonumber \\
+|\alpha_{me}|^{2}(\frac{1}{\epsilon^2} {\flow}_{m}+\mu^2 {\flow}_{e})] \nonumber\\
-\frac{16 \pi \mu}{c}[\frac{1}{\epsilon}\Im(\alpha_{e}^{*} \alpha_{me})+ \mu\Im(\alpha_{m}^{*}\alpha_{me})]<{\bf S}>\nonumber \\
+2[\frac{1}{\epsilon}\Re(\alpha_{e}^{*} \alpha_{me})- \mu\Re(\alpha_{m}^{*}\alpha_{me})]
\Im({\bf E} \times{\bf B}^{*}) \} .  \label{tscatt11}
\ee
$<w>$ denotes the density of electromagnetic energy: $<w>=<w_e>+<w_m>$.

It is interesting that if the particle, in addition to  being chiral, is dual \cite{kerk,suppre,greffin}:  $\epsilon^{-1}\alpha_{e}=\mu \alpha_{m}$,  then    (\ref{tordip311}) and (\ref {tscatt11})   become
\be
 {\cal L}' = \frac{n}{ k  } \{\frac{\alpha_{e}^{R}}{\epsilon} k_0  \nabla{\hel}
+\frac{2\alpha_{e}^{I}}{\epsilon}[\frac{ k^2}{c}{\flow} 
-  \frac{2\pi }{c}\nabla \times <{\bf S}>] \nonumber \\
-4\pi\frac{1}{\epsilon} \alpha_{me}^{I} \nabla <w> 
-\frac{nk}{c}\frac{1}{\epsilon}  \alpha_{me}^{R}\nabla \times  {\flow}  \nonumber \\
+  \frac{8\pi k }{c} \sqrt{\frac {\mu}{\epsilon}}\alpha_{me}^{R}<{\bf S}> \}.  ,\label{tordip41}
\ee
\be
 {\cal L}^{s} = -\frac{2 k^3}{3 c} \{
 nk[\frac{1}{\epsilon^2}|\alpha_{e}|^{2} {\flow}
+|\alpha_{me}|^{2}(\frac{1}{\epsilon^2} {\flow}_{m}+\mu^2 {\flow}_{e})] \nonumber\\
-8 \pi \frac{\mu}{\epsilon}\Im(\alpha_{e}^{*} \alpha_{me})<{\bf S}>\}.\,\
  \label{tscatt21}
\ee
Thus the reactive parts dissapear. We obtain two kind of terms in (\ref{tordip41}) and (\ref{tscatt21}): those conservative in ${\cal L}'$ linked to the gradient of the helicity and energy, and those non conservative  in ${\cal L}'$ and  ${\cal L}^s$ containing the flows of helicity and energy, as well as their curls. Once again, this has a suggestive analogy with the gradient, radiation pressure and curl components  of the optical force \cite{MNV2010}.

\section{The  orbital angular momentum}
In the non-absorbing dielectric medium of  refractive index $n=\sqrt{\epsilon \mu}$
 the  field density of orbital angular momentum  ${\bf L}$ and of  its flow  $\bm\Lambda_{ij}$, ($i,j=1,2,3$),  are defined  in the dual-symmetric formulation here employed, as \cite{bliokh2}:
\begin{eqnarray}
{\bf L}=\frac{c}{2 \mu}[{\cal E} \cdot {\leh} {\cal A}+ {\cal B} \cdot {\leh} {\cal C}],    \label{f1} 
\\ 
{\leh}={\bf r} \times \nabla. \label{L}
\end{eqnarray}
\be
 {\bm\Lambda}_{ij}=\frac{c^2}{2\mu}[\epsilon_{ikl}\epsilon_{jmn}r_k({\cal B}_n  \partial_l {\cal A}_m - {\cal E}_n  \partial_l {\cal C}_m) \nonumber \\ 
+{\cal B}_j {\cal A}_i - {\cal E}_j {\cal C}_i]    , \,\,\,\,\,
(i,j,k,l,m,n=1,2,3). \label{tens11}
\ee
 $\epsilon_{ikl}$ is the antisymmetric Levi-Civita tensor. In (\ref{f1}) and (\ref{tens11})   ${\bf \cal A}$ and ${\bf \cal C}$ are the vector potentials introduced in Section 2.

Under the premises established in Section 2, ${\bf L}$ and ${\bm\Lambda}_{ij}$ fulfill the continuity equation \cite{bliokh2}
\begin{eqnarray}
\dot{\bf L} + \nabla \cdot  {\bm\Lambda}_{ij} =- {\MM} \label{contL1}.
\end{eqnarray}
Where $-{\MM}$ represents  the dissipation by transfer of orbital angular  momentum on interaction of the fields with matter.

In addition to (\ref{potsr}), the time-harmonicity of the fields leads to the time-averages \cite{bliokh2}:
\begin{eqnarray}
{\bf L}=<{\bf L}>=  \frac{c}{4n k } \Im \{\epsilon {\bf E}^*  \cdot  {\leh} {\bf E} + \frac{1}{\mu}{\bf  B}^* \cdot {\leh}{\bf  B}\},    \label{f} 
\ee
\be
 {\bm\Lambda}_{ij}=< {\bm\Lambda}_{ij}>=\frac{c^2}{4kn\mu} \Im \{\epsilon_{jkl}( B_{l } ^{*} {\leh} E_k +E_l  {\leh}  B_{k } ^{*}) \nonumber \\
+{ B}_{i }^{*} E_j +{ B}_{j }^{*} E_i  \} ,  \,\,\,\,\,\,\,\,
(i,j=1,2,3). \label{tens} 
\ee
Where   ${\MM}$ in (\ref{contL1})  is replaced by its average:
\begin{eqnarray}
<{\MM}>=\frac{\pi c}{nk} \{\Im [{\bf J}^{*} \cdot {\leh} {\bf E}] - \Re [{\bf J}^{*} \cdot {\leh} {\bf E}] \nonumber \\
 - \partial_j \Im[\epsilon_{ikl}\epsilon_{jmn} x_k  (\partial_l K_m)  E_{n}^{*} + E_{j}^{*} K_i]\}
 . \label{pp}
\end{eqnarray}
$x_k$ being the $kth$-Cartesian component of the position vector ${\bf r}$.

\section{Conservation of the orbital angular momentum in scattering}

Equation (\ref{contL1}) expresses the conservation of orbital angular momentum flow and, again, consveys a density of  force moment, or torque. We next find what does this mean. As before, we integrate it  on any volume  surrounding the scattering object. The divergence term is then transformed into a flow  across the volume surface $\Sigma$  that will eventually be taken as a sphere of large radius $r$. 

With reference to Fig.1 let us first consider a  monochromatic, elliptically polarized, plane wave incident on an arbitrary body. The field at any point of the surrounding medium is again represented as the sum of the incident  and the scattered field.  Then the  flow (or time-averaged flow) density of orbital angular momentum  may be written by means of  ${\bm\Lambda}_{ij}$ decomposed as: ${\bm\Lambda}_{ij}={\bm\Lambda}_{ij}^{(i )}+ {\bm\Lambda}_{ij}^{(s)}+{\bm\Lambda}_{ij}' $ .  Where
\begin{eqnarray}
 {\bm\Lambda}_{ij}^{(i) }=\frac{c^2}{4kn\mu} \Im \{\epsilon_{jkl}( B_{l }^{(i) \,\,*} {\leh} E_k^{(i) } +E_l ^{(i) } {\leh}  B_{k } ^{(i) \,\,*}) \nonumber \\
+{ B}_{i }^{(i ) \,\,*} E_j ^{(i)} +{ B}_{j }^{(i) \,\,*} E_i ^{(i)} \} . \label{tens1} \\
 {\bm\Lambda}_{ij}^{(s)}=\frac{c^2}{4kn\mu} \Im \{\epsilon_{jkl}( B_{l } ^{(s)\,\,*} {\leh} E_k^{(s)} +E_l^{(s)}  {\leh}  B_{k } ^{(s)\,\,*}) \nonumber \\
+{ B}_{i }^{(s)\,\,*} E_j^{(s)} +{ B}_{j }^{(s)\,\,*} E_i^{(s)}  \}. \label{tens2} \\
 {\bm\Lambda}'_{ij}=\frac{c^2}{4kn\mu} \Im \{\epsilon_{jkl}( B_{l } ^{(i)\,\,*} {\leh} E_k ^{(s)}+E_l ^{(i)} {\leh}  B_{k } ^{(s)\,\,*} \nonumber \\
+ B_{l } ^{(s)\,\,*} {\leh} E_k^{(i )}+E_l^{(s)}  {\leh} + B_{k } ^{(i)\,\,*}) \nonumber \\
+{ B}_{i }^{i\,\,*} E_j^{(s)} +{ B}_{j }^{(i)\,\,*} E_i^{(s)}  +{ B}_{i }^{(s)\,\,*} E_j^{(i)} +{ B}_{j }^{(s)\,\,*} E_i ^{(i)} \}. \label{tens3} 
\ee
From Eq.(\ref{contL1}) the rate ${\Phi}_{\leh}$ at which the total orbital flow varies   on interaction of the wavefield with the body, is  given as before by the outward flux:  $-(n^2/ 4 \pi c^2)\int_{\Sigma}  d\Omega r^2 {\bm\Lambda}_{ij} \cdot {\bf s}$   through the surface of any large sphere $\Sigma$ of radius $r$ with center at some point $ {\bf r}_0$ of the region  occupied by the object. I.e.,  according to Eq.(\ref{contL1}):
\begin{equation}
{\Phi}_{\leh}=  {\Phi}_{\leh}^{(i)}+{\Phi}_{\leh} ^{(s)}+{\Phi}_{\leh}'. \label{Lho}
\end{equation}
Where ${\Phi}_{\leh}^{(i)}$, ${\Phi}_{\leh} ^{(s)}$ and  ${\Phi}_{\leh}'$ are  the $\Sigma$-integrals of the projections on $\bf s$ of  $ -{\bm\Lambda}_{ij} ^{(i)}$, $- {\bm\Lambda}_{ij} ^{(s)}$ and $- {\bm\Lambda}_{ij} ' $, respectively.  ${\Phi}_{\leh}=(n^2/ 4 \pi c^2)\int_{\Sigma} d r d\Omega r^2 <{\MM}>$.

From these equations we have that  ${\Phi}_{\leh} ^{(i)}=0$, so that  (\ref{Lho}) becomes
\be
{\Phi}_{\leh}={\Phi}_{\leh}^{(s)}+{\Phi}_{\leh}'. \label{lh}
\ee
Operating with  (\ref{tens1})-(\ref{tens3}) and using the identity: $\epsilon_{jlk}\epsilon_{kmn}= \delta_{jm}\delta_{ln}-\delta_{jn}\delta_{lm}$, the integrals  of $-{\bm\Lambda}_{ij}^{(s)} \cdot {\bf s}$ and $-{\bm\Lambda}_{ij}' \cdot {\bf s}$ across $\Sigma$  lead to:
\be
{\Phi}_{\leh}^{s}=-\frac{n^2}{ 4 \pi c^2}\int_{\Sigma}  d\Omega r^2  {\bm\Lambda}_{ij}^{(s)} s_j=\nonumber \\
 -\frac{1}{8\pi k}\int_{\Sigma}  d\Omega r^2  \Im[\epsilon  {\bf e}^{*}({\bf s}) \cdot  {\leh} {\bf e} ({\bf s})=\nonumber \\ 
-\frac{1}{8\pi k}\int_{\Sigma}  d\Omega r^2  \Im[\frac{1}{\mu} {\bf b}^{ *}({\bf s})\cdot  {\leh} {\bf b}({\bf s})=\nonumber \\ 
-\frac{1}{16\pi k}\int_{\Sigma}  d\Omega r^2  \Im[ \epsilon{\bf e}^{ *}({\bf s})\cdot  {\leh} {\bf e}({\bf s})+ \,\,\,\,\,\,\,\,\,\nonumber \\ 
\frac{1}{\mu} {\bf b}^{  *}({\bf s})\cdot  {\leh} {\bf b}({\bf s})]. \,\,\,\,\,\,\, \label{tss}
\ee
And using   again Jones' lemma, Eq. (\ref{psp1}), we get:
\be
{\Phi}_{\leh}'=-\frac{n^2}{ 4 \pi c^2}\int_{\Sigma}  d\Omega r^2    {\bm\Lambda}_{ij}'    s_{j}=\nonumber \\
-\frac{ 1 }{4 k^2 } \Re \{\epsilon{\bf e}^{(i)*}  \cdot [ {\leh}  {\bf e}({\bf s})]_{{\bf s}={\bf s}_{i} } \}= \nonumber \\
-\frac{ 1 }{4  k^2 } \Re \{\frac{1}{\mu}{\bf b}^{(i) *}  \cdot [ {\leh}  {\bf b}({\bf s})]_{{\bf s}={\bf s}_{i} } \} . \label{fit}
\ee
Eqs. (\ref{tss}) and  (\ref{fit}) constitute the  conservation law for the  orbital angular momentum of  either the electric or magnetic field, as well as of the dual-symmetric one with both ${\bf e}$ and ${\bf b}$.    Eq.(\ref{lh}), together with (\ref{tss})  and (\ref{fit}) define the force moment, or torque, by transfer of this orbital angular  momentum from the incident field to the object, expressed in terms of the flow of the tensor ${\bm\Lambda}_{ij}$ across any closed surface surrounding this body. It is very interesting that the rate at which this incident orbital angular  momentum ${\Phi}_{\leh}'$,  Eq.(\ref{fit}), is extinguished by transference to the body is proportional to a sort of cross-orbital angular momentum given by the interaction of the incident field with the scattered field in the forward direction $s = s_i$.

Moreover, recalling that  the incident field is  a plane wave, and since ${\bf r} = r{\bf s}$,  one has for the incident electric orbital angular momentum
\be
\Im[{\bf E}^{(i)*}  \cdot  {\leh} {\bf E}^{(i)} ]=\Im \{{\bf E}^{(i) *} ikr  {\leh} [{\bf  s }_i \cdot {\bf s}]{\bf E}^{(i)} \}= \nonumber \\
\Im \{{\bf e}^{(i) *} ikr  [{\bf  s } \times  {\bf s}_i]{\bf e}^{(i)} \}=|{\bf e}^{(i)}|^{2} kr  ({\bf  s } \times  {\bf s}_i). \,\,\,\,\,\, \label{OAMin}
\ee
With an analogous expression for the magnetic  momentum. Then the total  incident orbital angular momentum enclosed in a sphere of large radius $R$ is s given by 
\be
\int_{0}^{R} dr r^2 \int_{0}^{2\pi} d\phi\int_{0}^{\pi} d\theta sin \theta\Im[{\bf E}^{(i) *}  \cdot  {\leh} {\bf E}^{(i)} +{\bf B}^{(i) *}  \cdot  {\leh} {\bf B}^{(i)} ] \nonumber \\
=0. \,\,\,\,\nonumber \,\,
\ee
As expected.   Therefore the extinction torque from  this orbital angular momentum, Eq. (\ref{fit}), is zero if the incident wave is plane. (However, as we shall see next, in general it is not zero for any arbitrary field).

 On the other hand, the physical meaning of the flow of $-{\bm\Lambda}_{ij}^{s}$, Eq.(\ref{tss}), corresponds to a force moment of the scattered field on the particle. This is the consequence for this recoil torque  of the conservation of orbital angular momentum. But since since the incident plane wave has no orbital moentum, this recoil contribution has to come from a transfer of the incident spin angular momentum into the scattered orbital angular momentum. Hence this effect constitutes the essence of the {\it spin-orbit} interaction \cite{bliokh1,bliokh2} as a consequence of the scattering of the incident field by the object.

\section{Effects on a magnetodielectric bi-isotropic dipolar particle: The electromagnetic torque involved in the conservation of the orbital angular momentum}
\subsection{Incident plane wave}
Let us now address a magnetodielectic bi-isotropic dipolar  particle. Using Eqs.(\ref{consti1})-(\ref{dipm1}), evaluating the angular integrals of (\ref{tss}), and substituting the results in (\ref{lh}), we obtain the extinction torque from the orbital angular momentum transfer to this scattering body:
\be
{\Phi}_{\leh}'=-
\frac{n }{4 \mu  }
 \Re [(\frac{1}{\epsilon} (  {\bf p} \cdot {\bf s}_{i}) {\bf B}^{(i)*}   - \mu   ( {\bf m} \cdot {\bf s}_{i}){\bf E}^{(i)*} ] .  \label{fiitt}
\ee
Likewise, we get the  scattering, or recoil, torque from the orbital angular momentum:
\be
{\Phi}_{\leh}^{s}=-\frac{k^3}{6} \Im \{\frac{1}{\epsilon} {\bf p}^{*}  \times {\bf p} + \mu  {\bf m}^{*}  \times {\bf m} \}. \label{ttss}
\ee
Eq.(\ref{ttss}) is identical to (\ref{ttss1}) for the scattering torque from the spin. Both recoil torques sum to yield the resultant scattering torque, which coincides with that derived from the conservation of   the total angular  momentum ${\bf J}$ through the integration: $\int_{\Sigma} r{\bf s} \times T_{ij} s_j ds$, \cite{nieto2015}. 

On the other hand, Eqs. (\ref{consti1}) and the transversality of the incident field show that ${\Phi}_{\leh}'$  given by  (\ref {fiitt}) is zero, as expected from the above discussion. However Eq.(\ref{ttss}) exhibits the non-zero scattering contribution ${\Phi}_{\leh}^s$ (recoil torque) to ${\Phi}_{\leh}$, Eq.(\ref{lh}), which  as mentioned before  expresses  the {\it spin-orbit  interaction}  involving the conversion and transfer of the  spin    into orbital angular momentum torque by scattering with the  object.
\subsection{Generalization to an arbitrary incident field}

The above equations are generalized to an arbitrary illuminating wavefield expressible by its angular spectrum (\ref{ang111}), which introduced in (\ref{fiitt}) yields
\be
{\Phi}_{\leh}'= \frac{1 }{4 k} \sqrt{\frac{\epsilon}{ \mu}} \Im \{ \frac{1}{\epsilon} ({\bf p} \cdot\nabla){\bf B}^{*}-\mu ({\bf m} \cdot\nabla){\bf E}^{*} \} .\,\,\,\,  \label{torq3}
\ee
This form  is  equal to half the contribution from the spatial structure of the  incident  field   to the extinction torque derived above from the spin conservation, as seen  by comparing  (\ref{torq3}) with the second term of  (\ref{torq31}).

 On te other hand, Eq. (\ref {ttss}) for ${\Phi}_{\leh}^{s}$ remains valid for an arbitrary incident  field. Thus we obtain from (\ref{lh})  using  (\ref {ttss}) and (\ref{torq3})  the following torque involved in the conservation of orbital angular momentum:
\be
{\Phi}_{\leh} =  \frac{1 }{4 k} \sqrt{\frac{\epsilon}{ \mu}} \Im \{ \frac{1}{\epsilon} ({\bf p} \cdot\nabla){\bf B}^{*}-\mu ({\bf m} \cdot\nabla){\bf E}^{*} \}
  \nonumber \\
-\frac{k^3}{6}   \Im \{\frac{1}{\epsilon} {\bf p}^{*}  \times {\bf p} + \mu  {\bf m}^{*}  \times {\bf m} \}.  \label{torfin1}  
\ee

\section{Decomposition of the angular momentum torque into conservative and non-conservative components}
Using (\ref{consti1}) and  taking into account the  vector identities (\ref{vectident}),  we express the two components  ${\Phi}_{\leh}'$, (\ref{torq3}), and  ${\Phi}_{\leh}^{s}$, (\ref{ttss}), of the orbital conservation  torque,  (\ref{torfin1}), of the transversal fields, again droping the superindex $(i)$, as
\be
 {\Phi}_{\leh}' = \frac{n}{2 k  \mu} \{\alpha_{e}^{R} [k_0\frac{\mu}{\epsilon}\nabla{\hel} -
\frac{1}{2\epsilon} \nabla \times \Im({\bf E} \times{\bf B}^{*})]\nonumber \\
+\alpha_{m}^{R} [k_0\mu^2\nabla{\hel} +
\frac{\mu}{2} \nabla \times \Im({\bf E} \times{\bf B}^{*})]\nonumber \\
+\frac{\alpha_{e}^{I}}{2\epsilon}[\frac{4\mu k^2}{c}({\flow}_m - {\flow}_e)
-  \frac{8\pi \mu}{c}\nabla \times <{\bf S}>+ \nabla  \Re({\bf E} \cdot{\bf B}^{*})] \nonumber \\
+\frac{\mu \alpha_{m}^{I}}{2} [\frac{4\mu k^2}{c}({\flow}_m - {\flow}_e)
-  \frac{8\pi \mu}{c}\nabla \times <{\bf S}>- \nabla  \Re({\bf E} \cdot{\bf B}^{*})] \nonumber \\
-8\pi\frac{\mu}{\epsilon}( \alpha_{me}^{I} \nabla <w_e>- \alpha_{em}^{I} \nabla  <w_m> )\nonumber \\
-\frac{2nk}{c}\frac{\mu}{\epsilon} ( \alpha_{me}^{R}\nabla \times  {\flow}_{e} -\alpha_{em}^{R}\nabla \times  {\flow}_{m} ) \nonumber \\
+\frac{k\mu}{n}(\alpha_{em}^{I}+\alpha_{me}^{I}) \Im({\bf E} \times{\bf B}^{*})
\}, \,\,\,\, \label{tordip2}
\ee
and
\be
 {\Phi}_{\leh}^{s} = -\frac{k^3}{6}
 \{\frac{4nk}{c}[\frac{1}{\epsilon^2}|\alpha_{e}|^{2} {\flow}_{e}+\mu^2 |\alpha_{m}|^{2} {\flow}_{m}\nonumber \\
+\frac{1}{\epsilon^2}|\alpha_{em}|^{2} {\flow}_{m}+\mu^2 |\alpha_{me}|^{2} {\flow}_{e}] \nonumber\\
+\frac{16 \pi \mu}{c}[\frac{1}{\epsilon}\Im(\alpha_{e}^{*} \alpha_{em})- \mu\Im(\alpha_{m}^{*}\alpha_{me})]<{\bf S}>\nonumber \\
-2[\frac{1}{\epsilon}\Re(\alpha_{e}^{*} \alpha_{em})+ \mu\Re(\alpha_{m}^{*}\alpha_{me})]
 \} .  \label{tscatt}
\ee
If the particle is chiral, $\alpha_{em}= -\alpha_{me}$,  (\ref{tordip2}) and  (\ref{tscatt}) reduce to
\be
 {\Phi}_{\leh}' =  \frac{n}{k  \mu} \{\alpha_{e}^{R} [k_0\frac{\mu}{\epsilon}\nabla{\hel} -
\frac{1}{2\epsilon} \nabla \times \Im({\bf E} \times{\bf B}^{*})]\nonumber \\
+\alpha_{m}^{R} [k_0\mu^2\nabla{\hel} +
\frac{\mu}{2} \nabla \times \Im({\bf E} \times{\bf B}^{*})]\nonumber \\
+\frac{\alpha_{e}^{I}}{2\epsilon}[\frac{4\mu k^2}{c}({\flow}_m - {\flow}_e)
-  \frac{8\pi \mu}{c}\nabla \times <{\bf S}>+ \nabla  \Re({\bf E} \cdot{\bf B}^{*})] \nonumber \\
+\frac{\mu \alpha_{m}^{I}}{2} [\frac{4\mu k^2}{c}({\flow}_m - {\flow}_e)
-  \frac{8\pi \mu}{c}\nabla \times <{\bf S}>- \nabla  \Re({\bf E} \cdot{\bf B}^{*})] \nonumber \\
-8\pi\frac{\mu}{\epsilon} \alpha_{me}^{I} \nabla <w> 
-\frac{2nk}{c}\frac{\mu}{\epsilon}  \alpha_{me}^{R}\nabla \times  {\flow} \}. \,\,\,\,\,\,\, \label{tordip3}
\ee
\be
 {\Phi}_{\leh}^{s} = - \frac{k^3}{6}
 \{\frac{4nk}{c}[\frac{1}{\epsilon^2}|\alpha_{e}|^{2} {\flow}_{e}+\mu^2 |\alpha_{m}|^{2} {\flow}_{m}\nonumber \\
+|\alpha_{me}|^{2}(\frac{1}{\epsilon^2} {\flow}_{m}+\mu^2 {\flow}_{e})] \nonumber\\
-\frac{16 \pi \mu}{c}[\frac{1}{\epsilon}\Im(\alpha_{e}^{*} \alpha_{me})+ \mu\Im(\alpha_{m}^{*}\alpha_{me})]<{\bf S}>\nonumber \\
+2[\frac{1}{\epsilon}\Re(\alpha_{e}^{*} \alpha_{me})- \mu\Re(\alpha_{m}^{*}\alpha_{me})]
\Im({\bf E} \times{\bf B}^{*}) \} .  \label{tscatt1}
\ee
And when the particle is dual, ($\alpha_{e}/\epsilon=\mu\alpha_{m}$), Eqs. (\ref{tordip3}) and (\ref {tscatt1})   become
\be
 {\Phi}_{\leh}' = \frac{n}{ k  } \{\frac{\alpha_{e}^{R}}{\epsilon} k_0  \nabla{\hel}\nonumber \\
+\frac{2\alpha_{e}^{I}}{\epsilon}[\frac{ k^2}{c}({\flow}_m - {\flow}_e) 
-  \frac{2\pi }{c}\nabla \times <{\bf S}>] \nonumber \\
-4\pi\frac{1}{\epsilon} \alpha_{me}^{I} \nabla <w> 
-\frac{nk}{c}\frac{1}{\epsilon}  \alpha_{me}^{R}\nabla \times  {\flow}  \}.  \label{tordip4}
\ee
\be
 {\Phi}_{\leh}^{s} = -\frac{2k^3}{3c} \{
 nk[\frac{1}{\epsilon^2}|\alpha_{e}|^{2} {\flow}\nonumber \\
+|\alpha_{me}|^{2}(\frac{1}{\epsilon^2} {\flow}_{m}+\mu^2 {\flow}_{e})] \nonumber\\
-8 \pi \frac{\mu}{\epsilon}\Im(\alpha_{e}^{*} \alpha_{me})<{\bf S}>\}.
  \label{tscatt2}
\ee
Similar remarks as for the analogous decomposition of the torque from the spin,  above, apply to these latter equations.

\section{The total electromagnetic torque}
From all the previous analysis we conclude that  the {\it total electromagnetic torque} on the particle,  $<{\bm \Gamma}>$, stemming from the conservation laws of both the spin and  orbital angular momenta, is given by the sum of ${\cal L}$, Eq.(\ref{lh1}),  [cf. also (\ref{ttss1}) and (\ref{lprimfin1}) or (\ref{torq31})] and ${\Phi}_{\leh}$, Eq.(\ref{torfin1}), which we express as:  
\be
<{\bm \Gamma}>=<{\bm \Gamma}'>+<{\bm \Gamma}^{s}>. \label{torqresu}
\ee
Where the  electromagnetic {\it extinction torque} is
\be
<{\bm \Gamma}'>={\cal L'}+{\Phi'}_{\leh}=  \,\,\,\,\,\,\,\,\,\,\,\,\,\,\,\,\,\,   \nonumber \\
<{\bf \Gamma}_0> 
+\frac{3}{4k} \sqrt{\frac{\epsilon}{\mu}}\Im\{\frac{1}{\epsilon}({\bf p} \cdot\nabla){\bf B}^{(i) *}-\mu  ({\bf m} \cdot\nabla){\bf E}^{(i) *} \}=\nonumber \\
- \frac{<{\bf \Gamma}_0>}{2}\
 +\frac{3}{4k} \sqrt{\frac{\epsilon}{\mu}}\Im\{
\frac{1}{\epsilon}p_j \partial_i B_{j}^{(i) *}- \mu m_j \partial_i E_{j}^{(i) *}\}. \,\,\,\,\,\,\,\label{torqresuprim}
\ee
And the electromagnetic {\it recoil} or {\it scattering torque} reads:
\be
<{\bm \Gamma}^{s}>={\cal L}^s+{\Phi}_{\leh}^s= -\frac{k^3}{3} \Im \{\frac{1}{\epsilon} {\bf p}^{*}  \times {\bf p} + \mu {\bf m}^{*}  \times {\bf m} \}. \,\,\, \label{torqrecoi}
\ee
As mentioned above, we emphasize that this total torque is also obtained from the conservation of the total angular momentum $<{\bf J}>$ \cite{nieto2015}. The analysis given here, however, elucidates the contribution of both the spin and the orbital angular momenta through their respective conservation laws, and shows that both torques are added, like the angular momenta.

We notice that making use of (\ref{consti1}) one has, (dropping the superindex $ (i)$ in all following equations, understanding that we deal with incident fields),
\be
<{\bm \Gamma}_0>+<{\bm \Gamma}^s>=[\frac{\alpha_{e}^{I}}{2}-\frac{k^3}{3}(\frac{|\alpha_{e}|^{2}}{\epsilon} \nonumber \\
+\mu |\alpha_{me}|^{2})] \Im ({\bf E}^{*} \times {\bf E}) \nonumber \\ 
+[\frac{\alpha_{m}^{I}}{2}-\frac{k^3}{3}(\mu |\alpha_{m}|^{2}+\frac{ |\alpha_{em}|^{2}}{\epsilon})] \Im ({\bf B}^{*} \times {\bf B}) \nonumber \\ 
+[\frac{\alpha_{me}^{R}-\alpha_{em}^{R}}{2}-\frac{2 k^3}{3}(\frac{1}{\epsilon} \Im(\alpha_{e}^{*}\alpha_{em}) \nonumber \\
+\mu \Im(\alpha_{m}\alpha_{me}^{*})
] \Re ({\bf E} \times {\bf B}^{*}) \nonumber \\ 
-[\frac{\alpha_{me}^{I}+\alpha_{em}^{I}}{2}-\frac{2 k^3}{3}(\frac{1}{\epsilon} \Re(\alpha_{e}^{*}\alpha_{em}) \nonumber \\
+\mu \Re(\alpha_{m}\alpha_{me}^{*})
] \Im ({\bf E} \times {\bf B}^{*}) .  \,\,\,\,\label{topt3}
\ee
However, introducing (\ref{consti1}) in the optical theorem that represents the conservation of energy \cite{MNV2010}
\be
 {\cal W}^{a}+\frac{c k^4}{3 n} [\epsilon^{-1} |{\bf p}|^2 + \mu |{\bf m}|^{2}]= 
\frac{\omega}{2} \Im[ {\bf p}  \cdot {\bf e}_{i}^{*}   +{\bf m}\cdot  {\bf b}_{i}^{*}] ; \,\,\,\label{top}
\ee
$ {\cal W}^{a}$ standing for the rate of energy absorption by the particle, one  has
\be
\frac{ {\cal W}^{a}}{\omega}=  [\frac{\alpha_{e}^{I}}{2}-\frac{k^3}{3}(\frac{|\alpha_{e}|^{2}}{\epsilon}+\mu |\alpha_{me}|^{2})] |{\bf E}|^{2}  \nonumber \\ 
+[\frac{\alpha_{m}^{I}}{2}-\frac{k^3}{3}(\mu |\alpha_{m}|^{2}+\frac{ |\alpha_{em}|^{2}}{\epsilon})] |{\bf B}|^{2} \nonumber \\ 
+[\frac{\alpha_{me}^{R}-\alpha_{em}^{R}}{2}-\frac{2 k^3}{3}(\frac{1}{\epsilon} \Im(\alpha_{e}^{*}\alpha_{em}) \nonumber \\
+\mu \Im(\alpha_{m}\alpha_{me}^{*})
] \Im ({\bf E} \cdot {\bf B}^{*}) \nonumber \\ 
+[\frac{\alpha_{me}^{I}+\alpha_{em}^{I}}{2}-\frac{2 k^3}{3}(\frac{1}{\epsilon} \Re(\alpha_{e}^{*}\alpha_{em}) \nonumber \\
+\mu \Re(\alpha_{m}\alpha_{me}^{*})
] \Re ({\bf E} \cdot {\bf B}^{*})
 . \,\,\,\,\,\,\,\,\,\,\,\label{top2}
\ee
It is interesting that for fields such that ${\bf  B}=\mp {\bf E}$, (like e.g. the circularly polarized field scattered from a dual particle, mentioned  in Section 4), one has from (\ref {topt3}) and (\ref {top2}) that
\be
<{\bm \Gamma}_0>+<{\bm \Gamma}^s>= \frac{\epsilon \sigma^{(a)}}{8\pi k}\Im({\bf E}^{(*)} \times {\bf E})=  \frac{n \sigma^{(a)}}{2\pi c} {\flow}_e.  \label{torquecirc}
\ee
Where we have defined the absorption cross-section as: $ \sigma^{(a)} =(8 \pi k /\omega |{\bf E}|^2 ) {\cal W}^{(a)}  $. This occurs in particular for  an incident circularly polarized plane wave and for some beams, as well as for a pure electric, or   a pure magnetic, dipolar particle, [in this latter case ${\flow}_e$ should be replaced by ${\flow}_m$ in (\ref{torquecirc})]. In such situations we see in (\ref{torquecirc}) that the component $<{\bm \Gamma}_0>$ of the extinction torque  $<{\bm \Gamma}'>$,[cf.(\ref{torqresuprim})], is cancelled by $<{\bm \Gamma}^s>$, thus only remaining in the sum  (\ref{torquecirc}) that part of  $<{\bm \Gamma}^s>$  which contains $\sigma^{(a)}$.

More generally, for any wavefield,  if the particle is magnetodielectric, although not bi-isotropic, ($\alpha_{em}=\alpha_{me}=0$), separating in (\ref {top2}) the electric and magnetic parts by writing $ {\cal W}^{a}= {\cal W}_{e}^{a} + {\cal W}_{m}^{a}$ and introducing the electric and magnetic cross-sections:  $ \sigma_{e}^{(a)} =(8 \pi k /\omega |{\bf E}|^2 ) {\cal W}_{e}^{(a)}  $ and  $ \sigma_{m}^{(a)} =(8 \pi k /\omega |{\bf B}|^2 ) {\cal W}_{m}^{(a)}  $, one has on writing the optical theorem   (\ref {top2}) separately for the electric and the magnetic parts,
\be
<{\bm \Gamma}_0>+<{\bm \Gamma}^s>= \frac{\epsilon} {8\pi k}\sigma_{e}^{(a)}\Im({\bf E}^{(*)} \times {\bf E}) \nonumber \\
 +\frac{\epsilon} {8\pi k}\sigma_{m}^{(a)}\Im({\bf B}^{(*)} \times {\bf B}) =  \frac{n }{2\pi c}( \sigma_{e}^{(a)}{\flow}_e +\sigma_{m}^{(a)}{\flow}_m) .  \label{torqueEB}
\ee
These equations are seen to be compatible with (\ref{tordip21}), (\ref{tscatt11}), (\ref{tordip2}) and  (\ref{tscatt}). 

Hence, in contrast with some previous work \cite{laporta,dunlop,chaumet2,dogariu,cana0,cana,beckeva} and in agreement with some experiments and calculations \cite{mars,chen1,brasse,garces,dunlop2,dunlop3,dunlop4}, these equations show that this particle does not  experiences a torque and spins due to the so far called intrinsic torque: $<{\bm \Gamma}_0>$, [which, as we have just seen, is cancelled by a part of the recoil torque on this kind of particles through the optical theorem relating imaginary parts and moduli of the permittivities, Eq.(\ref{top2})], but because it absorbes the incident energy and, as shown by (\ref{torquecirc}) and (\ref{torqueEB}),  receives  the spin ${\flow}_e$ or ${\flow}_m$ through the remaining part of the recoil torque that contains the absorption cross section. On the  other hand, as we  see below, the orbiting of the particle is due to the beam shape carried out in the torque by the second term of (\ref{torqresuprim}). 

In this regard, it is thus pertinent to remark that in contrast with  previous torque  theories that do not comply with the optical theorem, the proportionality factors between the optical torque and the electric and magnetic  spin angular momenta is given by the absorption cross sections, not by the imaginary part of the corresponding electric and magnetic polarizabilities. This is a consequence of the static starting point and hence of the lack of consideration  of the recoil torque  in such theories. 

It is well known that the static regime applies for Rayleigh particles, (for which  $k a << 1$, $a$ being their linear dimension, i.e. for example their radius if they are spheres) \cite{born,MNV2010,Nieto2011}. In this limit the optical torque from such theories coincides with the one obtained in this paper.

If for example the dipolar particle is purely dielectric, ($ \alpha_{m}=\alpha_{em}=\alpha_{me}=0$, $\sigma_{e}^{(a)}$ is just denoted as $\sigma^{(a)}$),  Eq.(\ref{top2}) which becomes   $-(2/3\epsilon) k^3 |\alpha_e|^2=-\alpha_{e}^{I}+ \epsilon\sigma^{(a)}/4\pi k$, introduced in the   recoil torque $<{\bm \Gamma}^s>=(1/3 \epsilon)|\alpha_{e}|^2 \Im[{\bf E}^{*} \times {\bf E}]$, [cf. Eq.(\ref{topt3})], leads to cancellation of $<{\bf \Gamma}_0>$ by the corresponding $\alpha_{e}^{I}$ term of $<{\bm \Gamma}^s>$. Hence, Eqs. (\ref{torqresu}), (\ref{torqresuprim}) and (\ref{topt3}) lead to a torque on this particle:
\be
<{\bm \Gamma}>=\frac{\epsilon\sigma^{(a)}}{8\pi k}\Im[{\bf E}^{ *} \times {\bf E}] 
+\frac{3}{4k} \sqrt{\frac{\epsilon}{\mu}}\Im\{\frac{1}{\epsilon}({\bf p} \cdot\nabla){\bf B}^{ *} \}. \,\,\,\,\,\label{torqconabs}
\ee
We see at once that the first term of  (\ref{torqconabs}), which we extend to right side of (\ref{torqueEB}), and that represents the {\it spin torque} that we denote as $<\bm \Gamma_{\sigma}>$, {\it characterizes  the rotation of the particle on its axis due to the spin transfer by absorption of the incident energy} through $\sigma^{(a)}$. 
 
In connection with the aforementioned static approximation, notice again that the above relationship:  $-(2/3\epsilon) k^3 |\alpha_e|^2=-\alpha_{e}^{I}+ \epsilon\sigma^{(a)}/4\pi k$ becomes:  $\alpha_{e}^{I}= \epsilon\sigma^{(a)}/4\pi k$ only in the Rayleigh limit \cite{born,MNV2010,Nieto2011}.

Concerning the orbital movement due to the field structure, generally involved in  the second term of  (\ref{torqresuprim}),  it is useful to consider the vast kind of wavefields that posses a vortex
\cite{allenlibro,allenlibro1,babiker,yao}. Expressing   ${\bf E}({\bf r})$ and ${\bf B}({\bf r})$, as well as the  position vector ${\bf r}$, in cylindrical coordinates $(R,\phi,z)$ and extracting the vortex phase  $e^{i l \phi}$, we write such incident fields as
\be
{\bf E}({\bf r})= e^{i l \phi} \tilde{{\bf E}}({\bf r}), \,\,\,\,\,\,\,\,\,
{\bf B}({\bf r})= e^{i l \phi} \tilde{{\bf B}}({\bf r}). \label{vortexfields}
\ee
$l$ being the topological charge. So that  from (\ref{vortexfields}) and  after Eq.(\ref{consti1}), we express this term of  (\ref{torqresuprim}) as
\be
\frac{3}{4k} \sqrt{\frac{\epsilon}{\mu}}\Im\{\frac{1}{\epsilon}({\bf p} \cdot\nabla){\bf B}^{ *}-\mu  ({\bf m} \cdot\nabla){\bf E}^{ *} \} \nonumber \\
=\frac{3}{4k} \sqrt{\frac{\epsilon}{\mu}} \{ (\frac{1}{\epsilon} \alpha_{e}^{R}- \mu \alpha_{me}^{R})[\Im(\tilde{{\bf E}} \cdot\nabla)\tilde{{\bf B}}^{ *}-
\frac{l}{R} \Re[\tilde{ E}_{\phi}\tilde{{\bf B}}^{ *} )] \nonumber \\
+(\frac{1}{\epsilon} \alpha_{em}^{R}- \mu \alpha_{m}^{R})[\Im(\tilde{{\bf B}} \cdot\nabla)\tilde{{\bf B}}^{ *} -
\frac{l}{R} \Re(\tilde{ B}_{\phi}\tilde{{\bf B}}^{ *} )] \nonumber \\
+(\frac{1}{\epsilon} \alpha_{e}^{I}- \mu \alpha_{me}^{I})[\Re(\tilde{{\bf E}} \cdot\nabla)\tilde{{\bf B}}^{ *} +
\frac{l}{R} \Im(\tilde{ E}_{\phi}\tilde{{\bf B}}^{ *} )]  \nonumber \\
+(\frac{1}{\epsilon} \alpha_{em}^{I}- \mu \alpha_{m}^{I})[\Re(\tilde{{\bf B}} \cdot\nabla)\tilde{{\bf B}}^{ *}+
\frac{l}{R} \Im(\tilde{ B}_{\phi}\tilde{{\bf B}}^{ *} )] \}. \,\,\,\,\,\, \label{lorbit}
\ee
For example, if the particle is purely dielectric and isotropic, (only $\alpha_e \neq 0$), then the right side of  (\ref{lorbit})  reduces to
\be
\frac{3}{4k} \sqrt{\frac{\epsilon}{\mu}}\Im\{\frac{1}{\epsilon}({\bf p} \cdot\nabla){\bf B}^{ *}  \nonumber \\
=\frac{3}{4kn}  \{ \alpha_{e}^{R}[\Im(\tilde{{\bf E}} \cdot\nabla)\tilde{{\bf B}}^{ *}-
\frac{l}{R} \Re[\tilde{ E}_{\phi}\tilde{{\bf B}}^{ *} )]\nonumber \\
+ \alpha_{e}^{I}[\Re(\tilde{{\bf E}} \cdot\nabla)\tilde{{\bf B}}^{ *} +
\frac{l}{R} \Im(\tilde{E}_{\phi}\tilde{{\bf B}}^{ *} )], \label{eleclorbit}
\ee
which  accounts for  {\it the particle   orbital movement}, and that expressing  again $\alpha_{e}^{I}$ in terms of $\sigma^{(a)}$ through the aforementioned  optical theorem:  $\alpha_{e}^{I}=\epsilon\sigma^{(a)}/4\pi k+(2/3\epsilon) k^3 |\alpha_e|^2$, shows that there is an $l$-number torque component,  proportional  to $l \sigma^{(a)}/R$, making the particle to orbit around the vortex axis via the  transfer of orbital angular momentum by absorption; also existing another orbital component by transfer through the  particle scattering cross section involved in $l |\alpha_e|^2/R$ \cite{MNV2010,suppre}. Both components are amenable of observation  \cite{allenlibro1}.

For fields with a well defined incidence direction, like optical beams, the term $(3/4kn) \alpha_{e}^{I}\Re(\tilde{{\bf E}} \cdot\nabla)\tilde{{\bf B}}^{ *}$ of (\ref{eleclorbit}), which  plays for the torque a role analogous to the $(1/2) \alpha_{e}^{I} \Im\{({\bf E} \cdot\nabla){\bf E}^{ *} \}$ component  of the orbital momentum in the optical force \cite{berry,MNV2010}, may make negative the resultant longitudinal   torque  (i.e. opposite to the incident  helicity). This is illustrated below with  incident Bessel and Gaussian beams.

 There are  more components  of this extrinsic torque, (\ref{eleclorbit}), in the transversal ${\bm\phi}$ and ${\bf R}$ directions; but for the purely dielectric   particle   the $\hat{\bm z}$ component of the two  terms of $\alpha_{e}^{R}$ in  (\ref{eleclorbit}) is zero for a variety of beams, like  those discussed in  the following examples. More complex effects of this sort remain to be studied in Eq.(\ref{lorbit}) when $ \alpha_{m}$, $\alpha_{em}$ and $\alpha_{me}$ are non-zero.

\section{Example 1: Incident Bessel beam}
As a first illustration we consider  a  transversal electric (TE) Bessel beam \cite{novits,yu} propagating along $OZ$ in air, with a  vortex of topological charge $l$, whose electric  and magnetic vectors are
\be
{\bf E}^{(i)}({\bf r})=\frac{e}{k} \, e^{il\phi} e^{ik_z z}(-\frac{lx}{R^2}J_l(KR)+i\frac{Ky}{R}J'_l(KR) , \,\,\,\,\,\,\,\,\nonumber \\
 -\frac{ly}{R^2}J_l(KR)-i\frac{Kx}{R}J'_l(KR) ,0); \,\,\,\,\,\,\,\,\,\,\nonumber \\
{\bf B}^{(i)}({\bf r})= \frac{ek_z}{k^2} \, e^{il\phi} e^{ik_z z}(\frac{ly}{R^2}J_l(KR)+i\frac{Kx}{R}J'_l(KR) , \,\,\,\,\,\,\,\,\nonumber \\
 -\frac{lx}{R^2}J_l(KR)+i\frac{Ky}{R}J'_l(KR) , \frac{K^2}{k_z}J_l(KR)); \,\,\,\,\,\,\,\,\,\,\label {bessf} \\
{\bf k}=(k_x,k_y,k_z),\,{\bf K}=(k_x,k_y,0); \,\,k=|{\bf k}|=\frac{\omega}{c}; \,\,\,\,\,\,\,\,  \nonumber\\
\epsilon=\mu=n=1;\,\,{\bf r}=(x,y,z), \,{\bf R}=(x,y,0), \,\,\,\,\,\,\,\,\nonumber
\ee
incident on a dielectric particle whose induced dipole moment is ${\bf p}=\alpha_e {\bf E}^{(i)}$.

This beam is eliptically polarized. First, we note that the helicity of this field is
\be
{\hel}=e^2\frac{l k_z K}{k^4 R}  J_{l}(KR)J'_{l}(KR) \label{beshel}
\ee

 Then according to (\ref{lprimfin1}) or (\ref{torq31}):
\be
{\cal L}'= 
\frac{1}{2 k} \Im p_j \partial_i B_{j}^{(i )*}=< {\bm  \Gamma}_0>
+\frac{1}{2 k} \Im \{({\bf p} \cdot \nabla){\bf B}^{(i) *}\} .\,\,\,\,  \label{btorq312}
\ee
With
\be
< {\bm  \Gamma}_0>= \alpha_{e}^{I}\,\frac{e^2}{k^2} \frac{lK}{R}J_l(KR)J'_l(KR) (0,0,1) \nonumber \\
=\frac{k^2}{k_z} \alpha_{e}^{I}{\hel} (0,0,1).  \label{gamma01}
\ee
Which shows that this component of the torque follows the helicity. Also, Eq.(\ref{gamma01})  manifests a contribution of  the vortex charge $l$ to this part of the torque. 

On the other hand,
\be
{\cal L}'=2 {\Phi}_{\leh}'=\frac{1}{2 k} \Im \{({\bf p} \cdot \nabla){\bf B}^{(i) *}\}=  \nonumber \\
\frac{ e^2 k_z}{2k^4} \{ \alpha_{e}^{R}
\frac{l K}{R^2}[K J_l(KR) J''_{l}(KR) +K J^{' 2}_{l}(KR) \,\,\,\,\,\,\,\,\,\,\,\,\,\,\,\nonumber \\
-\frac{1}{R} J_{l}(KR) J'_{l}(KR)](x,y,0) +\alpha_{e}^{I}[[\frac{l^2}{R^4} J^{2}_{l}(KR)\,\,\,\,\,\,\,\,\,\,\,\,\,\,\, \nonumber \\
-2\frac{l^2 K}{R^3} J_{l}(KR)J'_{l}(KR)  +\frac{K^2}{R^2} J^{' 2}_{l}(KR)](y,-x,0)  \,\,\,\,\,\,\,\,\,\,\,\,\,\,\, \nonumber \\
-2\frac{l K^3}{k_zR} J_{l}(KR)J'_{l}(KR)(0,0,1) ] \}
 .\,\,\,\,  \label{btorqpb}
\ee
Where $J'_{l}(KR)=d J_{l}(KR)/d(KR)$. 

Eq.(\ref{btorqpb}) exhibits an  azimuthal dependence $(y, -x) = -R \hat{\bm \phi}$  of the $\alpha_{e}^{I}$ terms.  ($R$ and $\phi$ are the cylindrical coordinates in the transversal plane of the beam).  Hence, although   $< {\bm  \Gamma}_0>$ is  longitudinal, i.e. along $OZ$, the beam structure creates and modulate a transversal azimuthal component of ${\cal L}'$. Moreover,  there is a  radial torque  $(x,y)=R \hat{\bm r}$ in the $\alpha_{e}^{R}$ terms  due to the helicity gradient. This is extensive to the torque part: $\frac{1}{4k} \Im \{({\bf p} \cdot \nabla){\bf B}^{(i) *}\}$ [cf. Eq.(\ref{torfin1})] coming from the orbital angular momentum  conservation law.

It is interesting to analyze the different parts in Eqs.(\ref{btorq312})-(\ref{btorqpb}) from the point of view of the decompositions    (\ref{tordip21}) and (\ref{tordip2}).  There are the conservative components: 
\be
\frac{1}{2}\alpha_{e}^{R}\nabla{\hel}=\frac{ e^2}{2k^4}\frac{l k_{z}K}{R^2}  \alpha_{e}^{R}
[K J_l(KR) J''_{l}(KR)  \nonumber \\
+K J^{' 2}_{l}(KR)-\frac{1}{R} J_{l}(KR) J'_{l}(KR)](x,y,0) \label{btorgrad}
\ee
\be
\frac{1}{4 k}\alpha_{e}^{I} \nabla  \Re({\bf E} \cdot{\bf B}^{*}) =0.   \,\,\, \label{btornab}
\ee
Eq.(\ref{btorgrad}) shows the aforementioned radial gradient torque whose sign oscillates with the distance $R$ to the beam axis.

The non-conservative components:
\be
\frac{k}{c}\alpha_{e}^{I} {\flow}_e = \frac{1}{2} < {\bm  \Gamma}_0>. \label{btorqin}
\ee
And 
\be
\frac{k}{c}\alpha_{e}^{I} {\flow}_m -\frac{2\pi}{ k c}\alpha_{e}^{I}\nabla \times <{\bf S}>=
\frac{ e^2 k_z}{2k^4}\alpha_{e}^{I}[[\frac{l^2}{R^4} J^{2}_{l}(KR)   \,\,\,\,\,\,\,\,\,\,\,\,\,\,\,\,\,\,\,\,\,\,\,\, \nonumber \\
-2\frac{l^2 K}{R^3} J_{l}(KR)J'_{l}(KR)   
+\frac{K^2}{R^2} J^{' 2}_{l}(KR)](y,-x,0) \,\,\,\,\,\,\,\,\,  \nonumber \\
-2\frac{l K^3}{k_zR} J_{l}(KR)J'_{l}(KR)(0,0,1) ] .\,\,\,\,\,\,\,\,\,\,\,\,\,\,\, \label{btorS}
\ee
Having used: $x^2 J''_l(x)+xJ'_l(x)+(x^2-l^2)J_l(x)=0$.

Also,
\be
-\frac{1}{4 k  }\alpha_{e}^{I} \nabla \times \Im({\bf E} \times{\bf B}^{*}) =0. \,\,\,\, \label{btorim} 
\ee
On the other hand, 
\be
 {\cal L}^s =  {\Phi}_{\leh}^{s}=-\frac{ k}{3} e^2 |\alpha_{e}|^2 \frac{lK}{R}J_l(KR)J'_l(KR) (0,0,1)
 ;\,\,\,\,  \label{btorqsca1}
\ee
which is contributed by the electric helicity flow term: $ -\frac{k^3}{6}
 \frac{4nk}{c}[\frac{1}{\epsilon^2}|\alpha_{e}|^{2} {\flow}_{e}$ of Eqs.(\ref{tscatt11}) and (\ref{tscatt}).

Using as shown above the optical theorem: $\alpha_{e}^{I}-(2/3\epsilon) k^3 |\alpha_e|^2= \epsilon\sigma^{(a)}/4\pi k$,  and  since $K^2 +k_{z}^{2}=k^2$, the resulting torque becomes
\be
<{\bm\Gamma}({\bf r})>=\frac{e^2}{4\pi k^3}\sigma^{(a)}\frac{lK}{R} J_{l}(KR) J'_{l}(KR)](0,0,1)\nonumber \\
+\frac{ 3e^2 k_z}{4k^4} \{ \alpha_{e}^{R}
\frac{l K}{R^2}[K J_l(KR) J''_{l}(KR)  +K J^{' 2}_{l}(KR)  \,\,\,\,\,\,\,\,\,\,\,\,\,\,\,\nonumber \\
-\frac{1}{R} J_{l}(KR) J'_{l}(KR)](x,y,0) 
+\alpha_{e}^{I}[[\frac{l^2}{R^4} J^{2}_{l}(KR)\,\,\,\,\,\,\,\,\,\,\,\,\,\,\, \nonumber \\
-2\frac{l^2 K}{R^3} J_{l}(KR)J'_{l}(KR)   
+\frac{K^2}{R^2} J^{' 2}_{l}(KR)](y,-x,0)\,\,\,\,\,\,\,\,\,\,\,\,\,\,\, \nonumber \\
-2\frac{l K^3}{k_zR} J_{l}(KR)J'_{l}(KR)(0,0,1) ] \}.\,\,\,\,\,\,\,\,\,\,\,\,\,\,\,
 \,\,\,\,  \label{btorgausspar2}
\ee
The spin torque $<{\bm\Gamma}_{\sigma}>$, [see Eq.(\ref{torqconabs})], that results from the sum:  $<{\bm\Gamma}_0>+{\cal L}^{s}+{\Phi}_{\leh}^{s}$, [cf. Eqs. (\ref{btorqin}) and  (\ref{btorqsca1})], is the first term   of (\ref{btorgausspar2}), and as said above,  describes the particle spinning following the incident helicity by transfer of  SAM through energy absorption. Eq.(\ref{btorgausspar2}) also  shows that, in particular if the wave were plane,  this term would be the only contribution to the torque,   there being  no orbiting of the particle,  as expected.   Otherwise,  there is a radial helicity-gradient term of $\alpha_{e}^{R}$, and an azimuthal vortex term of  $\alpha_{e}^{I}$ in (\ref{btorgausspar2}) due to both the beam transversal structure and  the interaction of the  transversal $\bf E$ and $\bf B$ with the longitudinal $\bf B$. The latter manifested through the particle  scattering and absorption of the incident beam as seen from the optical theorem, and  producing an  orbiting of the particle around the vortex axis $OZ$ at   maximum intensity points, which are  stable due to the gradient force. 

\begin{figure}[htbp]
\centerline{\includegraphics[width=1\columnwidth,height=8cm]{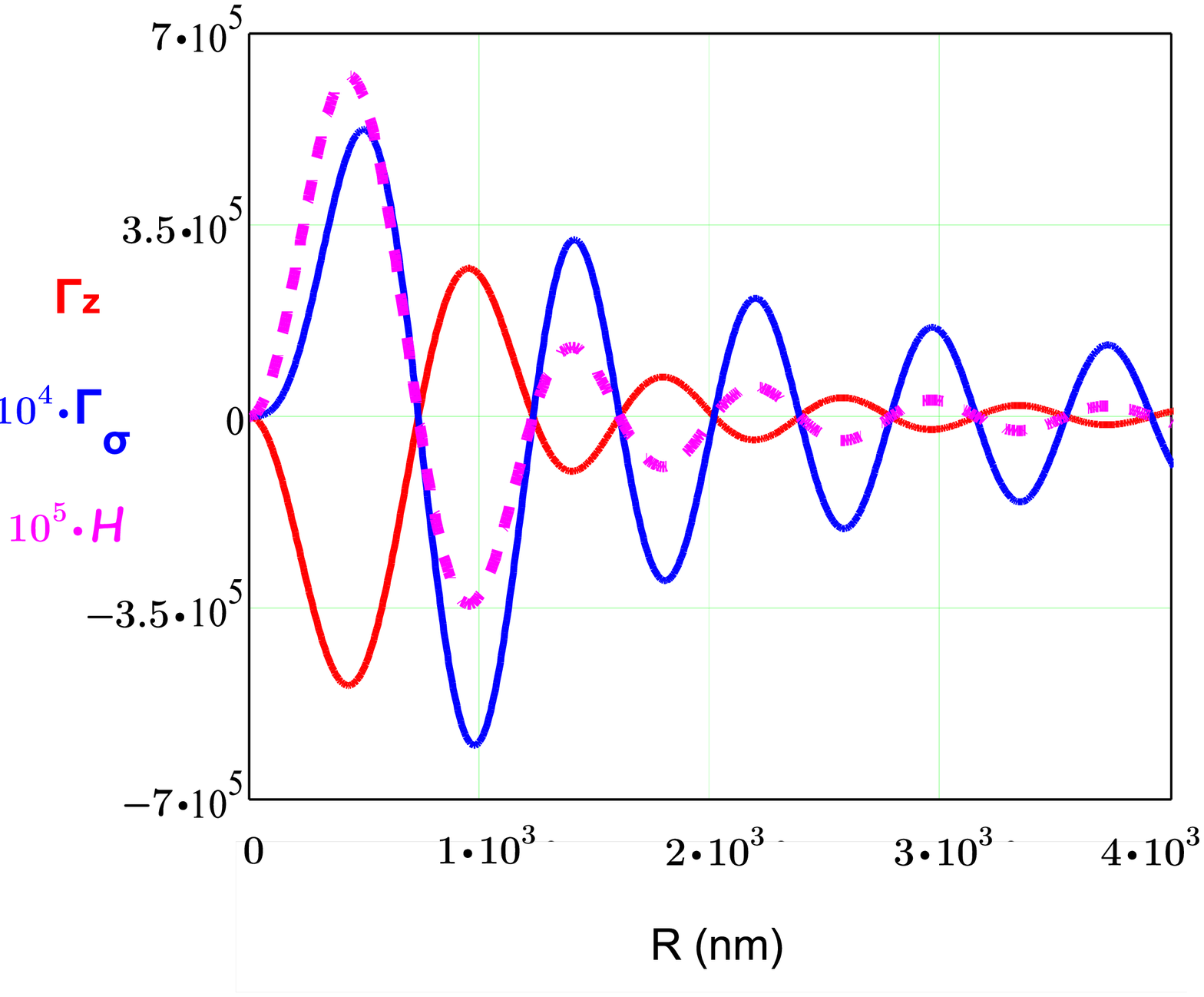}}
\caption{(Color online). Doped  $Si$ sphere of radius $230$ nm illuminated at $\lambda=1350$ nm, where its electric polarizability dominates, by a TE Bessel beam.  $K=0.9k$, $l=2$,  $\alpha_{e}^{R} =  1.095\times 10^7$nm$^3$, $\alpha_{e}^{I}=  6.083\times 10^6$nm$^3$, $\sigma^{(a)}=0.1$.  Red: Longitudinal component of the time-averaged optical torque $<{\bm \Gamma}>_z$. Blue: Longitudinal spin torque $<\bm \Gamma_{\sigma}>$ transferred by absorption, (its magnitude appears multiplied by $10^4$).  Broken pink: Helicity   ${\cal H}$ of the incident beam, (its magnitude appears multiplied by $10^5$). The three quantities have circular symmetry with annular spatial distribution in the transversal $XY$-plane, and thus in this representation they are even functions  if  the axis of the radial coordinate  $R$  is extended to the left of $0$ . The torques and helicity  are plotted  in nm$^3$ and nm$^{-1}$, respectively, since they are normalized to half the incident intensity density: $e^2$.}
\end{figure}
Fig. 2 shows three quantities in the transversal plane, where they have an annular spatial distribution, for a Si sphere, (refractive index $n_p = 3.5$) illuminated at a wavelength near its dipolar resonance \cite{Nieto2011,luki}, and $K=0.9k$. The particle is assumed to have been slightly doped to confer it a low absorption, so that its spin torque is observable  through  $\sigma^{(a)}$. The optical torque $<\bm \Gamma>_{z}$  is dominated by the last term of  (\ref{btorgausspar2}), which is several orders of magnitude larger than the spin torque $<\bm \Gamma_{\sigma}>$. Also, while the latter follows the incident field helicity, the former opposes it, thus giving rise to a longitudinal  $\bf z$-negative optical torque on the particle, which consequently orbits around the vortex axis in opposite sense to its spinning.  Diminishing $K$ results in a weaker torque, but does not change its negative sign. Similarly occurs on varying $l$, which only changes the oscillation points along $R$, slowly diminishing their ampliitudes as $|l|$ grows. 

It is also interesting in connection with the gradient torque (\ref{btorgrad}) and the corresponding behavior of  the helicity  in Fig. 2, that this optical torque transversal component is maximum  and zero at those points where   ${\hel}$ changes sign and is extreme, respectively. These positions coincide with those where the gradient force magnitude on the particle  is maximum and zero, respectively. Hence while this latter force tends to place the object at an equilibrium position of minimum energy, the gradient torque tends to a zero transversal value, both in its spinning and orbital movement, (cf. Eqs.  (\ref{tordip21}) and (\ref{tordip2}) for the spin and angular gradient optical torques). As anticipated in Section V, this effect  can be exploited as an additional degree of freedom in what  we may call  a {\it rotational optical tweezer}.

\section{Example 2: Incident Gaussian beam}
\subsection{TE and TM Gaussian beam}
We shall now consider a  wavefield commonly used in several works and illustrations. This is a   Gaussian beam in  air, both TE and TM,  circularly polarized, with  spot size $2\sigma$ and a  vortex of topological charge $l$: 
\be
{\bf E}^{(i)}({\bf r})=e(1,\pm i, 0) e^{il\phi} e^{ikz}e^{-\frac{R^2}{\sigma^2}}.\nonumber \\
{\bf B}^{(i)}({\bf r})=\frac{1}{ik}\nabla \times {\bf E}^{(i)}({\bf r}) \simeq \frac{1}{ik} \partial_z \times {\bf E}^{(i)}({\bf r}) =\hat{\bf z}\times {\bf E}^{(i)}({\bf r})\nonumber \\
=\mp i {\bf E}^{(i)}({\bf r})=e (\mp i, 1,0) e^{il\phi}e^{ikz}e^{-\frac{R^2}{\sigma^2}},  \,\,\,\,\,\,\,\,\,\,\,\, \label{pgaussf} \\
{\bf k}=(k_x,k_y,k_z),\,{\bf K}=(k_x,k_y,0); \,\,k=|{\bf k}|=\frac{\omega}{c}; \,\,\,\,\,\,\,\,  \nonumber\\
\epsilon=\mu=n=1;\,\,{\bf r}=(x,y,z), \,{\bf R}=(x,y,0), \,\,\,\,\,\,\,\,\nonumber
\ee
illuminating a dielectric particle whose induced dipole moment is ${\bf p}=\alpha_e {\bf E}^{(i)}$. 

This is an interesting  case because as we see below, although some aspects of the forces stemming from the above paraxial expressions of the $ {\bf E}$ and $ {\bf B}$ vectors are easily obtained, these fields are not divergenceless and, as we  see next,  are not appropriate for any theory  based on the free-space  Maxwell equations: $\nabla \cdot {\bf E}=\nabla \cdot {\bf B}=0$, like the torque formulation developed here. Neither the total  angular momentum density, nor its $\bf z$-flux across a transversal $XY$-plane, can be expressed as the sum of the corresponding spin plus orbital densities and fluxes,  as expected from the remarks of Section II. 

First, we note that the helicity of this field is
\be
{\hel}=\pm\frac{1}{k} e^2e^{-\frac{2R^2}{\sigma^2}}. \label{phel}
\ee
It is convenient to analyze  the dynamics exerted by this beam on a dielectric particle from the point of view of the mean optical force, which is \cite{arias,albala,berry,MNV2010}
\be
<{\bf F}>=\frac{\alpha_{e}^{R}}{4} \nabla |{\bf E}|^2+ \alpha_{e}^{I}\frac{4\pi k}{c}{\bf P}_0 . \,\,\,\,\, \label{force}
\ee
We omit the  superindex ${(i)}$, understanding that in all following equations we are dealing with the incident field.

The gradient, pulling the particle to the beam axis, is
\be
\nabla |{\bf E}|^2=\frac{4}{\sigma^2}e^2 e^{-\frac{2R^2}{\sigma^2}}(-x,-y,0). \label{grad} \\ (-x,-y,0)=-R \hat{\bf r}. \nonumber
\ee
The {\it ith} component of the orbital momentum ${\bf P}_0$ being \cite{berry}
\be
{\bf P}_0=\frac{c}{8\pi k}\Im\{ E_{j}^{*} \partial_i { E}_{j}\}=<{\bf S}>+\frac{c}{8\pi k}\Im\{({\bf E}^* \cdot \nabla){\bf E}\} .   \,\,\,\,\,   \label{orbi} \\
i,j=1,2,3. \nonumber
\ee
With the mean energy flow:
\be
<{\bf S}>=\frac{c}{8\pi}e^2 e^{-\frac{2R^2}{\sigma^2}} (0,0,1). \label{pspin}
\ee
And 
\be
\Im\{({\bf E}^* \cdot \nabla){\bf E}=e^2 e^{-\frac{2R^2}{\sigma^2}}(\pm\frac{2}{\sigma^2}-\frac{l}{R^2})(y,-x,0) . \label{porbit}
\ee
Now observe that since for the paraxial TE and TM field (\ref{pgaussf}), $\nabla \cdot {\bf E}\neq 0$, one has
\be
\Im\{({\bf E}^* \cdot \nabla){\bf E}=\Im\{(\nabla\cdot{\bf E}){\bf E}^*\} -\frac{1}{2}\nabla \times\Im\{{\bf E}^* \times {\bf E}\}. \,\,\,\,\,\label{spinculr}
\ee
The existence of the non-zero first term in the left side of (\ref{spinculr}) is a fact that should be recalled when expressing the orbital momentum in terms of the electric spin $<\flow_{e}>$ curl: $\nabla \times\Im\{{\bf E}^* \times {\bf E}\}$ for fields that are not divergenceless. 

Notice that the orbital momentum ${\bf P}_0$, described by (\ref{orbi})-(\ref{porbit}), contains the term $\pm\frac{2}{\sigma^2}$, [cf.Eq. (\ref{porbit})], which is essential to describe the spin curl component of the force, [cf.Eq. (\ref{spinculr})]:
\be
\frac{1}{2}\nabla \times\Im\{{\bf E}^* \times {\bf E}\}=\mp e^2 \frac{4}{\sigma^2} e^{-\frac{2R^2}{\sigma^2}} (y,-x,0). \label{curlspin} \\
(y,-x,0)= -R\hat{\bm \phi}.  \nonumber
\ee
Which together with 
\be
\Im\{(\nabla\cdot{\bf E}){\bf E}^*\}=e^2 e^{-\frac{2R^2}{\sigma^2}}(\mp\frac{2}{\sigma^2}-\frac{l}{R^2})(y,-x,0), \label{pdiv}
\ee
yield the expression  (\ref{porbit}) when they both are introduced in  (\ref{spinculr}). In fact, the omission of the $\pm\frac{2}{\sigma^2}$ leads to the well-known fundamental paradox of having a zero spin momentum, [the last term of Eq. (\ref{spinculr})] while the spin angular momentum $\flow_e$ is not zero \cite{pad}.

Eqs.(\ref{force}), and (\ref{orbi})-(\ref{porbit}), characterize the force due to ${\bf P}_0$, pushing the particle along $z$ with the radiation pressure characterized by $<{\bf S}>$, and in the azimuthal direction due to $\Im\{({\bf E}^* \cdot \nabla){\bf E}$ given by the divergence and spin curl terms  of (\ref{spinculr}). Also, from the above equations one has that the total angular momentum  $\bf J$ does not hold  $\bf J=\bf L+\bm\flow$, where $\bf J= \bf r \times <{\bf S}>$ and $\bf L= \bf r \times {\bf P}_0$. Neither  their corresponding flows across the transversal  $XY$-plane  fulfill such decomposition. 

One would write the torque ${\bf R}\times \alpha_{e}^{I}<{\bf S}>$,   given by the moment of the radiation pressure with respect to the beam axis, as azimuthal: $-\frac{c}{8\pi}e^2 e^{-\frac{2R^2}{\sigma^2}}  \alpha_{e}^{I}\hat{\bm \phi}$. Whereas the  torque  ${\bf R}\times \frac{\alpha_{e}^{I}}{2}\Im\{({\bf E}^* \cdot \nabla){\bf E}$ from  the moment of the remaining  orbital part of the force would be longitudinal: $-e^2 e^{-\frac{2R^2}{\sigma^2}}  \alpha_{e}^{I}R^2 (\pm\frac{2}{\sigma^2}-\frac{l}{R^2})\hat{\bm z}$, (notice this orbital $\hat{\bf z}$-component being proportional to the vortex index $l$ and independent of $R$), to which the moment of the spin contribution ${\bf R}\times \frac{\alpha_{e}^{I}}{2}(\nabla \times\Im\{{\bf E}^* \times {\bf E}\})$ is $\pm e^2 \frac{4}{\sigma^2} e^{-\frac{2R^2}{\sigma^2}} \alpha_{e}^{I}R^2\hat{\bm z} $.  However, as we have already seen, the   torque by transfer of  spin to the particle actually comes from its absorption cross section. If there is no absorption, (or birefringence \cite{garces}), there is no spinning of the particle due to the SAM.

On the other hand, according to (\ref{lprimfin1}) or (\ref{torq31}):
\be
{\cal L}'=\frac{n}{2 k \mu} \Im \frac{1}{\epsilon} p_j \partial_i B_{j}^{(i )*}=
< {\bm  \Gamma}_0> \nonumber \\
+\frac{1}{2 k} \sqrt{\frac{\epsilon}{\mu}} \Im \{\frac{1}{\epsilon}({\bf p} \cdot \nabla){\bf B}^{(i) *}\}=< {\bm  \Gamma}_0> +  2{\Phi}_{\leh}'  .\,\,\,\,  \label{ptorq312}
\ee
With  $\epsilon=\mu=n=1$, $ k=|{\bf k}|=\frac{\omega}{c}$, and

\be
< {\bm  \Gamma}_0>=\pm \alpha_{e}^{I}e^2  e^{-\frac{2R^2}{\sigma^2}} (0,0,1) \nonumber \\
= k\alpha_{e}^{I}{\hel} (0,0,1).\label{pgamma0}
\ee
\be
\frac{1}{2 k} \sqrt{\frac{\epsilon}{\mu}} \Im \{\frac{1}{\epsilon}({\bf p} \cdot \nabla){\bf B}^{(i) *}\} \nonumber \\
=  \frac{ e^2}{2k } e^{-\frac{2R^2}{\sigma^2}} \{ \alpha_{e}^{R}(\pm\frac{2}{\sigma^2}-\frac{l}{R^2})(-x,-y,0)\nonumber \\
\pm\alpha_{e}^{I}(\pm\frac{2}{\sigma^2}-\frac{l}{R^2})(y, -x, 0) \} .\,\,\,  \label{ptorqpb} \\
(\epsilon=\mu=n=1,  k=|{\bf k}|=\frac{\omega}{c}). \,\,\,\,\,\,\,\,\,\,\,\,\,\,\, \nonumber
\ee

As expected, Eqs.(\ref{ptorq312})-(\ref{ptorqpb})  show features similar to those seen above for the Bessel beam  concerning the influence of the structure of the beam, in both the longitudinal and tranversal components of these torque parts.

On the other hand the recoil torques are
\be
 {\cal L}^s =  {\Phi}_{\leh}^{s}=\mp\frac{ k^3}{3\epsilon} e^2 |\alpha_{e}|^2e^{-\frac{2R^2}{\sigma^2}} (0,0,1)
 ;\,\,\,\,  \label{ptorqsca1}
\ee
which are contributed by the electric helicity flow term: $ -\frac{k^3}{6}
 \frac{4nk}{c}[\frac{1}{\epsilon^2}|\alpha_{e}|^{2} {\flow}_{e}$ of Eqs.(\ref{tscatt11}) and (\ref{tscatt}).

When  the resultant torque, Eq.(\ref{torqresu}),  $<{\bm\Gamma}({\bf r})> ={\cal L}+{\Phi}_{\leh} ={\cal L}'+{\cal L}^{s}+{\Phi}_{\leh}'+{\Phi}_{\leh}^{s}$ exerted by this beam on the particle, is evaluated, one obtains:
\be
<{\bm\Gamma}({\bf r})>=e^2  e^{-\frac{2R^2}{\sigma^2}} \{\pm \frac{\epsilon}{4\pi k} \sigma^{(a)}(0,0,1) \nonumber \\ 
+\frac{3}{4k}[ \alpha_{e}^{R}(\pm\frac{2}{\sigma^2}-\frac{l}{R^2})(-x,-y,0) \nonumber \\
\pm\alpha_{e}^{I}(\pm\frac{2}{\sigma^2}-\frac{l}{R^2})(y, -x, 0)] \}. \,\, \,\,\,\,\,\,\,\,\,\,\,\label{ptorrgausspar}
\ee
Once again, we obtain in the first term of (\ref{ptorrgausspar}) the spin torque, with its $\pm$ sign, along $OZ$, proportional to the  absorption cross section, accounting for the particle spinning. We also see in  (\ref{ptorrgausspar}) the  helicity gradient  radial term, as well as an azimuthal part proportional to $\Im\{({\bf E}^* \cdot \nabla){\bf E}$ according to Eq.(\ref{porbit}).

No $z$-component due to OAM transfer is however exhibited by (\ref{ptorrgausspar}).  This  is a consequence of obviating the fundamental conditions: $\nabla\cdot \bf E=0$, $\nabla\cdot \bf B=0$ in the paraxial representation  (\ref{pgaussf})  of the Gaussian beam. Of course, neither all terms of the decompositions  (\ref {tordip21}) and (\ref {tordip2}) are valid for this beam. Hence, although some aspects of the force exerted by a  TE and TM beam, like that of (\ref{pgaussf}), are easily described with such representation of its $\bf E$ and $\bf B$ vectors,  other effects of the dynamics as discussed in Eq.(\ref{ptorrgausspar}), are not adecuately described.

\subsection{Gaussian beams satisfying the transversality condition}
 Paraxial expressions of vortex Gaussian  beams  that,  being   $\partial_z \simeq ik_z$, satisfy the transversality condition, should have  $E_{z}^{(i)}=(i/k_z) \nabla_{\perp} \cdot {\bf  E}_{\perp}^{(i)}$ \cite{berry},  ${\bf  E}_{\perp}^{(i)}= ( E_{x}^{(i)}, E_{y}^{(i)})$. A first example, that we denote as PG1, has the annular spatial field distribution in the $XY$-plane:  
\be
{\bf E}^{(i)}({\bf r})=ek^l R^l (1,\pm i , \frac{i}{k_z }e^{\pm i  \phi} [\frac{l}{R}(1 \mp 1) \nonumber \\
-\frac{2R}{\sigma^2}]) 
 e^{il\phi} e^{ik_z z}e^{-\frac{R^2}{\sigma^2}}.
\label{pg1}
\ee
Alternatively, a second instance that we denote as  PG2, is similar to an hypergeometric Gaussian beam in the  pupil plane:
\be
{\bf E}^{(i)}({\bf r})=e[1,\pm i, (\pm\frac{2}{k_z\sigma^2} \nonumber \\
+\frac{l}{k_zR^2})(y \mp ix)] e^{ik_z z}e^{-\frac{R^2}{\sigma^2}}.\,\, \,\,\,\,\,\,   \label {pg2} \\
{\bf B}^{(i)}({\bf r})=-i \frac{1}{k} \nabla\times{\bf E}^{(i)} .   \,\,\,\,\,\,\,\,\,\,\,\,\,\,\,\, \\
{\bf r}=(x,y,z), \,\,\,\,\,\,\, {\bf R}=(x,y,0). \,\,\,\,\,\,\,\,\,\,\,\,\,\,\,\,   \nonumber
\ee

\begin{figure}[htbp]
\centerline{\includegraphics[width=1.\columnwidth]{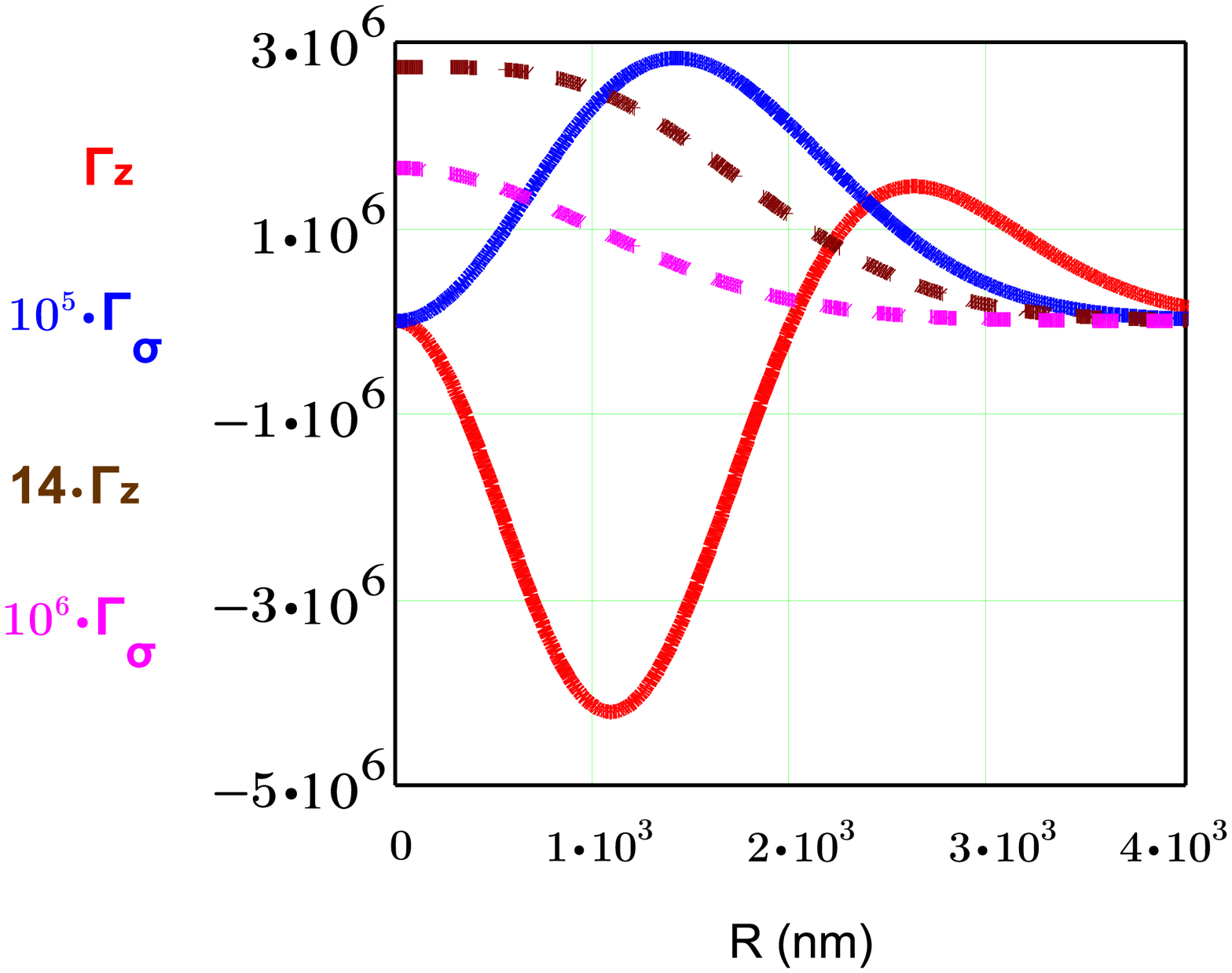}}
\caption{(Color online). Doped $Si$ sphere of radius $230 \,$nm illuminated at $\lambda=1350 \,$nm by either of the Gaussian beams  PG1 [Eq.(\ref{pg1})]  and PG2 [Eq.(\ref{pg2})] with LCP transversal components. $l=1$ for PG1, and $l=2$ for PG2.  $\sigma= 2 \mu m$. $\alpha_{e}^{R} =  1.095\times 10^7$nm$^3$, $\alpha_{e}^{I}=  6.083\times 10^6$nm$^3$. Full red: Longitudinal component of  $<\Gamma>_z$ due to PG1. Full blue: Longitudinal spin torque $<\Gamma_{\sigma}>$ from PG1, (its magnitude appears multiplied by $10^5$).   Broken brown: Longitudinal component of  $<\Gamma>_z$ due to PG2, (its magnitude appears multiplied by $14$). Broken pink: Longitudinal spin torque $<\Gamma_{\sigma}>$ from PG1, (its magnitude appears multiplied by $10^6$).  The four quantities have circular symmetry  with an  annular spatial distribution in the transversal $XY$-plane, and thus in this representation they are even functions  if  the axis of the radial coordinate  $R$  is extended to the left of $0$ . They  are plotted  in nm$^3$  since they are normalized to half the incident intensity density: $e^2$. }
\end{figure}

We choose $l=1$ and the upper signs in (\ref{pg1}) for the PG1 beam. According to the first term of (\ref{torqconabs}), the spin torque $<{\bm \Gamma}_{\sigma}>$ that it induces on the above dielectric dipolar particle is:
\be
<{\bm \Gamma}>_{\sigma}=e^2  k^2 R^2 e^{-\frac{2R^2}{\sigma^2}}\frac{\sigma^{(a)}}{2 \pi  k} \{\frac{1}{k_z \sigma^2}(-y,x,0) \nonumber \\
+\frac{1}{2}(0,0,1) \}.\,\,\,\,\,\,\,\,
\ee

Whereas by the same token, the PG2 beam gives rise to:
\be
<{\bm \Gamma}>_{\sigma}=\pm e^2  e^{-\frac{2R^2}{\sigma^2}}\frac{\sigma^{(a)}}{4 \pi  k} \{ (\pm\frac{2}{k_z\sigma^2}+\frac{l}{k_zR^2})(-y,x,0) \nonumber \\
+(0,0,1) \}.\,\,\,\,\,\,\,\,
\ee 
For brevity we concentrate on these longitudinal $z$- torques, as this component was the source of the discrepancy encountered  with (\ref{pgaussf}).    They are:
\be
<{\bm\Gamma}({\bf r})>_z=  e^{-\frac{2R^2}{\sigma^2}} R^2 [\frac{k \sigma^{(a)}}{4 \pi  }
\nonumber \\
+ 6\frac{e^2}{\sigma^2}\alpha_{e}^{I} (\frac{R^2}{\sigma^2}-1)] \}(0,0, 1),  \,\,\,\,\,\,\,\, \label{mtorgausspar1}
\ee
for the PG1 beam. And
\be
<{\bm\Gamma}({\bf r})>_z=e^2  e^{-\frac{2R^2}{\sigma^2}} \{\pm\frac{\sigma^{(a)}}{4 \pi  k}
\,\,\,\,\,\,\,\, \nonumber \\
+\frac{3}{k^2\sigma^2}(l\mp1\pm\frac{2R^2}{\sigma^2}) \alpha_{e}^{I} \}(0,0, 1) ,  \,\,\,\,\,\,\,\, \label{mtorgausspar2}
\ee
for the PG2 beam.

Eqs.(\ref{mtorgausspar1})  and (\ref{mtorgausspar2}) contain the topological charge, or orbital number, $l$, coming from the $<\bf S>$ component of the orbital momentum $\bf P_0$, Eq. (\ref{orbi}),  which describes the trajectory of the particle around the vortex  following the sign of  $l$. There is no contribution to this  $l$ term of   $<{\bm\Gamma}({\bf r})>_z$ from  $\Im [({\bf E}^{*} \cdot \nabla) {\bf E}]$ since, for example for PG2, the latter  yields  $4 e^2  e^{-\frac{2R^2}{\sigma^2}}(\pm 4/\sigma^2) (y,-x,0)$, whose momentum $\bf R \times \Im[(\bf E^* \cdot \nabla)\bf E] $ contributes to the $\mp  (3/k^2\sigma^2)$ factor of $\alpha_{e}^{I}$ in  (\ref{mtorgausspar2}). Of course, again  the sign of the incident spin angular momentum is followed by the  spinning particle  as it absorbes incident energy through ${\sigma^{(a)}}$. 

Fig. 3 shows the annular spatial distribution of the spin and  longitudinal total torques on the Si sphere addressed above, again  illuminated  at $\lambda=1350$ nm by either the beam PG1 (with $l=1$) or  PG2 (with $l=2$), choosing the  upper sign in (\ref{pg1})  and (\ref{pg2}). The longitudinal $OZ$-torque  is dominated by the $\alpha_{e}^{I}$ term of  Eqs.(\ref{mtorgausspar1})  and (\ref{mtorgausspar2}), which is several orders of magnitude larger than the spin torque $<\bm \Gamma_{\sigma}>$. The latter clearly displays the beam transversal  Gaussian  shape at whose peak the particle is stably placed by the gradient force, while it orbits around the vortex and spins on its axis.   Increasing  $|l|$  enlarges the strength of $<{\bm\Gamma}>_z$. 

It is remarkable that while the PG2 beam induces a resultant  $z$ torque that follows its helicity like the spin torque, the interaction of the transversal and longitudinal components of the PG1 wavefield induces a negative electromagnetic torque on the particle, which consequently orbits around the vortex axis in opposite sense to its spinning.  This example, as well as the one above concerning a TE Bessel beam, illustrate that in contrast with  negative radiation pressure and tractor beams \cite{chen,novits,dogariu1},  the production of negative torques is quite common and  exists beyond    illumination with special oscillating spatial profiles.

We should remark that in the examples here shown, the torque  is enhanced by illumination in  the electric dipole  resonance region of this kind of high $n_p$ particles. By contrast, we obtained  torques  one order of magnitude smaller  on latex spheres  ($n_p=1.5$) of the same size in the equivalent region ($\lambda=700$ nm).

\section{Conclusions}
We have studied the significance and consequences of the conservation laws of the spin and orbital angular momenta for the scattering of arbitrary wavefields, satisfying the transversality condition, by an object, which we have illustrated by a magnetodielectric bi-isotropic dipolar particle in the wide sense. We have shown  that these laws describe the respective contributions of these angular momenta to the different parts of the torque exerted by the field on the particle, which is seen to consist of an extinction and a scattering, or recoil,  component. The latter is shown to cancel the often called  intrinsic torque, contained in the extinction component and modelled in some previous works, without compatibility with energy conservation, to account for the  optical torque felt  by the body, and that as shown here plays no role in the resultant optical torque, which is actually felt by the particle through its absorption cross section. This latter quantity is what remains from the recoil torque component after the above mentioned cancellation.

Each of these two laws describe a half of the recoil  torque acting on the object,  The transfer of  spin and orbital angular momenta, on the other hand, arises while they are extinguished from the incident field. This is described by the torque extinction parts due to the interference of the incident and scattered fields. If the incident wave is plane, and thus it has no orbital angular momentum, its transfer characterized by the recoil torque is a manifestation of the spin-orbit interaction.

The role of the spatial structure of the incident field in the torque has been established, showing that the contribution from the spin angular momentum conservation  is twice that from the orbital angular momentum continuity law. While the spinning of the particle through absorption of the incident energy always follows the incident helicity, its orbiting  may result  opposite to the incident spin, this giving rise to a resultant negative optical torque on a single particle. I have illustrated this with Bessel and Gaussian beams.

The electromagnetic torque admits a decomposition into conservative and non-conservative components in which the helicity and its flow play a role analogous to that of the energy and the Poynting vector in the optical force. In particular, the gradient torque has the potential to add a new variable to optical tweezer set-ups.

In addition to contributing to an understanding of the physical mechanisms involved in the transfer of angular momentum from the field to a body, ruled by these conservation laws, and  their several consequences, this  study opens new avenues of research both from the experimental and theoretical points of view that, with the improvement of particle manipulation techniques, adds new degrees of freedom in the analysisis and control of possible fundamental phhenomena and their applications.

\section{Acknowledgments}
Work  supported by the MINECO through grants FIS2012-36113-C03-03 and FIS2014-55563-REDC.


\begin{thebibliography}{99}
\bibitem{allenlibro} L. Allen, S. M.  Barnett and M. J.  Padgett, eds, {\it Optical Angular Momentum}, (IOP Publishing, Bristol, UK, 2003).

\bibitem{allenlibro1} D. L. Andrews and M. Babiker, eds., {\it The Angular Momentum of  Light} (Cambridge U.P., Cambridge, 2013).

\bibitem{babiker} L. Allen, M. J. Padgett and M. Babiker, in {\it Prog. Opt.} {\bf 39}, E. Wolf, ed., (Elsevier, Amsterdam, 1999).

\bibitem{yao}M. Yao and M. Padgett, Adv. Opt. Photon. {\bf 3}, 161 (2011).

\bibitem{mansu}M. Mansuripur, Opt. Express {\bf  13},  5315 (2005).

\bibitem{barnett} S. M. Barnett, J. Opt. B: Quantum Semiclass. Opt. {\bf 4}, S7 (2002).

\bibitem{bliokh2}K. Y. Bliokh, J.Dressel and F. Nori, New J. Phys. {\bf 16},  093037 (2014). 

\bibitem{bliokh1}K. Y. Bliokh and F. Nori, Phys. Rep. {\bf  592}, 1  (2015).

\bibitem{marston1}J. H. Crichton and P. L. Marston, Electron. J. Dif. Eqs., {\bf Conf. 04}, 37 (2000). http://ejde.math.swt.edu or http://ejde.math.unt.edu 

J\bibitem{dola2}N. B. Simpson, K. Dholakia, L. Allen, and M. J. Padgett, Opt. Lett. {\bf 22}, 52 (1997).

\bibitem{lipkin} D.M. Lipkin,  J. Math. Phys. {\bf 5 },  696 (1964).

\bibitem{alex} C. N. Alexeyev,  Y. A. Friedman and  A. N.Alexeyev, Ukr. J. Phys. 2001. {\bf 46}, 43 (2001).

\bibitem{cameron1}R. P Cameron,  S.M Barnett and A. M Yao, New. J. Phys. {\bf 14}, 053050 (2012).

\bibitem{cameron2} R. P. Cameron and S. M. Barnett, New J. Phys. {\bf 14}, 123019 (2012).

\bibitem{philb}T. G. Philbin, Phys. Rev. A {\bf   87}, 043843 (2013).

\bibitem{tang1} Y. Tang and A. E. Cohen, Phys. Rev. Lett. {\bf  104}, 163901 (2010).

\bibitem{tang2} Y. Tang and A. E. Cohen, {\bf  332}, 333 (2011).

\bibitem{marru} L. Marrucci, E. Karimi, S.  Slussarenko, B. Piccirillo, E. Santamato, 
E. Nagali and F. Sciarrino,    J. Opt. {\bf 13},  064001 (2011).

\bibitem{molina1}J. P. Torres, G. Molina-Terriza and L. Torner, Proc. SPIE {\bf  5958} 51581O (2005).
\bibitem{molina2}I. Fernandez-Corbaton, X. Zambrana-Puyalto and G. Molina-Terriza, Phys. Rev. A {\bf 86}, 042103  (2012).

\bibitem{beth} R. A, Beth, Phys. Rev. {\bf 50},  115 (1936). 

\bibitem{garces} V. Garces-Chavez,  D. McGloin,  M. J. Padgett,  W. Dultz,  H. Schmitzer, and K. Dholakia,  Ph ys. Rev. Lett. {\bf  91}, 093602 (2003). 

\bibitem{grier}J. E. Curtis and D. G. Grier, Phys. Rev. Lett. {\bf 90}, 133901 (2003).

\bibitem{dunlop2} T. A. Nieminen, N. R. Heckenberg and H. Rubinztein-Dunlop, 
J. Mod. Opt. {\bf 48},  405 (2001).

\bibitem{dunlop3} T. Asavei, V. L. Y. Loke, N. R. Heckenberg and H. Rubinztein-Dunlop,  J. Quant. Spectr. Rad. Transf. {\bf 110}, 1472 (2009). 

\bibitem{dunlop4}M. E. J. Friese, T. A. Nieminen, N. R. Heckenberg, and H. Rubinsztein-Dunlop, Opt. Lett. {\bf 23}, 1 (1998). 

\bibitem{laporta}A. La Porta and M. D. Wang, Phys. Rev. Lett. {\bf 92}, 190801 (2004).

\bibitem{dunlop}A. Lehmuskero, P. Johansson, H. Rubinsztein-Dunlop, L. Tong, and M. Kall, ACS Nano  {\bf 9}, 3453 (2015).  (DOI: 10.1021/acsnano.5b00286).

\bibitem{chaumet2}P. C. Chaumet and C. Billaudeau, J. Appl. Phys. {\bf 101}, 023106 (2007).

\bibitem{dogariu}D. Haefner, S. Sukhov, and A. Dogariu, Phys. Rev. Lett. {\bf 103}, 173602 (2009).

\bibitem{cana0}A. Canaguier-Durand, A. Cuche, C. Genet and T. W. Ebbesen, Phys.  Rev. A {\bf 88}, 033831 (2013).

\bibitem{cana}A. Canaguier-Durand, J. A. Hutchison, C. Genet and T. W. Ebbesen, New J. Phys. {\bf 15}, 123037 (2013).

\bibitem{beckeva} K.Y. Bliokh, A.Y. Bekshaev, F. Nori, Nature Comm. {\bf  5}, 3300 (2014).

\bibitem{mars}P. L. Marston and J. H. Crichton,, Phys. Rev. A {\bf 30}, 2508 (1984).

\bibitem{dog} C. Schwartz and A.  Dogariu, Opt. Express {\bf 18}, 8425 (2006).

\bibitem{chen1}J. Chen, J. Ng, K. Ding, K. H. Fung, Z.Lin and C. T. Chan, Phys. Rep. {\bf 4}, 6386 (2014).

\bibitem{nieto2015} M. Nieto-Vesperinas, Opt. Lett  {\bf 40},  3021 (2015).

\bibitem{brasse} D. Hakobyan and E. Brasselet, Nature Photon.   DOI: 10.1038/NPHOTON.2014.142 (2014).

\bibitem{chen}J. Chen, J. Ng, Z. Lin, and C. T. Chan, Nat. Photonics 5, 531–534 (2011).

\bibitem{novits} A. Novitsky, C. W. Qiu, and H. Wang, Phys. Rev. Lett. 107, 203601 (2011).

\bibitem{dogariu1} S. Sukhov and A. Dogariu, Phys. Rev. Lett. 107, 203602 (2011).

\bibitem{Ashkin1} A. Ashkin, Phys. Rev. Lett. {\bf 24}, 156 (1970).

\bibitem{Ashkin2} A. Ashkin, J.M. Dziedzic, J. E. Bjorkholm, and S. Chu, Opt. Lett. {\bf 11}, 288 (1986); A. Ashkin, Proc. Natl. Acad. Sci. USA {\bf 94}, 4853 (1997); Science {\bf 210}, 1081 (1980).

\bibitem{grier} D.G. Grier, Nature {\bf 424}, 810 (2003).

\bibitem{neu} K.C. Neuman and S.M. Block, Rev. Sci. Instrum. {\bf 75}, 2787 (2004).

\bibitem{mazilu} M. Dienerowitz, M. Mazilu, and K. Dholakia, J. Nanophoton. {\bf 2}, 021875 (2008).

\bibitem{chaumet}P. C. Chaumet and M. Nieto-Vesperinas, Opt. Lett. {\bf 25}, 1065–1067 (2000).

\bibitem{arias}J. R. Arias-Gonzalez and M. Nieto-Vesperinas, J. Opt. Soc. Am. A {\bf 20}, 1201–1209 (2003).

\bibitem{quid1}  R. Quidant  and  C. Girard,   Laser and Photon. Rev. {\bf 2},  47–57 (2008).

\bibitem{quid2} M. L. Juan, R. Gordon, Y. Pang,  F. Eftekhari and R. Quidant, Nat. Phys. {\bf 5}, 915 (2009). 

\bibitem{albala} Silvia Albaladejo, M. I. Marques, M. Laroche, and J. J. Saenz, Phys. Rev. Lett. {\bf 102}, 1136021  (2009).

\bibitem{berry}M.V. Berry, J. Opt. A {\bf 11}, 094001- 094012 (2009).

\bibitem{chaumet3} P.C. Chaumet and A. Rahmani, Opt. Express {\bf 17}, 2224 (2009).

\bibitem{MNV2010} M. Nieto-Vesperinas, J. J. Saenz, R. Gomez-Medina, and L.
Chantada, Opt. Express{\bf  18}, 11428–11443 (2010).

\bibitem{nietoPRA1} M. Nieto-Vesperinas, Phys Rev. A {\bf 92}, 023813    (2015).

\bibitem{kong} J.A. Kong, Proc IEEE {\bf 60}, 1036 (1972).

\bibitem{Nieto2011} A. Garcia-Etxarri, R. Gomez-Medina, L. S. Froufe-Perez, C.
Lopez, F. Scheffold, J. Aizpurua, M. Nieto-Vesperinas, and J. J. Saenz, Opt. Express {\bf 19}, 4815 (2011).

\bibitem{chanlat} S.B. Wang and C.T. Chan, Nat. Comm. {\bf 5: 3307}, 4307 (2014).

\bibitem{andrews1}  M. M. Coles and D. L. Andrews, Opt.Lett. {\bf 37}, 3009 (2013).

\bibitem{andrews2} D. L. Andrews and M. M. Coles, Opt.Lett. {\bf 38}, 869 (2012).

\bibitem{andrews3} D. L. Andrews, M. M. Coles, M. D. Williams and D. S. Bradshaw, Proc. SPIE {\bf  8813}, 88130Y (2013).

\bibitem{pu}  Y. Li, H. Liu, Z. Chen, J. Pu and B. Yao,  Opt. Rev. {\bf 18}, 7 (2011).

\bibitem{boyd2} M. N. O'Sullivan, M. Mirhosseini, M. Malik and R. W.  Boyd, Opt. Express {\bf 20}, 24444 (2012).

\bibitem{nori} K. Y.  Bliokh, A. Y. Bekshaev and F. Nori, New J. Phys. {\bf 15}, 033026  (2013).

\bibitem{sherman} G.C. Sherman,  Phys. Rev. Lett. {\bf 21}, 761-764 (1968).

\bibitem{nietolib}M. Nieto-Vesperinas, Scattering and Diffraction in Physical Optics, (2nd edition, World Scientific, Singapore, 2006).

\bibitem{mandel}L. Mandel and E. Wolf, Optical Coherence and Quantum Optics, (Cambridge U.P., Cambridge, 1995).

\bibitem{jackson}J. D. Jackson, Classical Electrodynamics, (3rd edition, John Wiley, New York, 1998).

\bibitem{jones} D.S. Jones, Proc. Camb. Phil. Soc. {\bf 48}, 736 (1952).

\bibitem{born} M. Born and E. Wolf, {\it Principles of Optics}, 7 th edition, Cambridge U.P., Cambridge, 1999. 

\bibitem{kerk} M. Kerker, D. S. Wang, and C. L. Giles, J. Opt. Soc. Am. 73, 765 (1983).

\bibitem{suppre}M. Nieto-Vesperinas, R. Gomez-Medina, and J. J. Saenz, J. Opt. Soc. Am. A 28, 54 (2011).

\bibitem{greffin} J.M. Geffrin, B. Garcia-Camara, R. Gomez-Medina, P. Albella, L. S. Froufe-Perez, C. Eyraud, A. Litman, R. Vaillon, F. Gonzalez, M. Nieto-Vesperinas, J.J. Saenz and F. Moreno, Nat. Commun. {\bf 3}, 1171 (2012).

\bibitem{MNV_PTRSL} M. Nieto-Vesperinas, P.C. Chaumet and A.  Rahmani, Phil. Trans. R. Soc. Lond. A {\bf 362}, 719-737 (2004).

\bibitem{schellmann} J. A. Schellman, Chem. Rev. {\bf 75 }, 323  (1975).

\bibitem{bliokhmagnetoel} K.Y. Bliokh, Y.S. Kivshar, F. Nori, Phys. Rev. Lett. {\bf 113}, 033601 (2014).

\bibitem{yu}Y. Z. Yu and W. B. Dou, Prog. Electromag. Res.  Lett., {\bf 5}, 57 (2008).

\bibitem{luki} A. B. Evlyukhin, C. Reinhardt, A. Seidel, B. S. Luk’yanchuk and B. N. Chichkov,  Phys. Rev. B {\bf 82}, 045404 (2010).

\bibitem{pad} L. Allen and M.J. Padgett, Am. J. Phys. {\bf 70}, 567 (2002).
\end{thebibliography}
\end{document}